\documentclass{article}

\usepackage{arxiv}

\usepackage[utf8]{inputenc} % allow utf-8 input
\usepackage[T1]{fontenc}    % use 8-bit T1 fonts
\usepackage{hyperref}       % hyperlinks
\usepackage{url}            % simple URL typesetting
\usepackage{booktabs}       % professional-quality tables
\usepackage{amsfonts}       % blackboard math symbols
\usepackage{nicefrac}       % compact symbols for 1/2, etc.
\usepackage{microtype}      % microtypography
\usepackage{lipsum}
\usepackage{graphicx}

\usepackage{amsmath}
\usepackage{setspace}
\usepackage{array}
\usepackage{lineno}
\usepackage{bm}
\usepackage{color}
\usepackage{multirow}
\usepackage{ulem}
\usepackage{amssymb}
\usepackage{lineno}

\usepackage{subfig}

\usepackage{algorithm,algorithmic}
\usepackage[algo2e]{algorithm2e}

\title{Generative modeling of seismic data using diffusion models and its application to multi-purpose posterior sampling for noisy inverse problems}

\author{
 Chuangji Meng \\
 The National Engineering Research Center for Offshore Oil and Gas Exploration\\
  School of Information and Communications Engineering\\
  Xi'an Jiaotong University\\
  Xi'an 710049, China \\
  \texttt{cjmeng@xjtu.edu.cn} \\
   \And
 Jinghuai Gao \\
  The National Engineering Research Center for Offshore Oil and Gas Exploration\\
   School of Information and Communications Engineering\\
  Xi'an Jiaotong University\\
  Xi'an 710049, China \\
  \texttt{jhgao@mail.xjtu.edu.cn} \\
  \And
 Wenting Shang \\
  School of Computer Science and Technology\\
  Xi'an University of Posts and Telecommunications\\
  Xi'an 710049, China \\
  \texttt{shangwt90@126.com} \\
    \And
  Yajun Tian \\
  The National Engineering Research Center for Offshore Oil and Gas Exploration\\ 
   School of Information and Communications Engineering\\
  Xi'an Jiaotong University\\
  Xi'an 710049, China \\
  \texttt{yajun.tian@xjtu.edu.cn} \\
    \And
  Hongling Chen \\
   The National Engineering Research Center for Offshore Oil and Gas Exploration\\
  School of Information and Communications Engineering\\
  Xi'an Jiaotong University\\
  Xi'an 710049, China \\
  \texttt{859311743@qq.com} \\
     \And
  Tieqiang Zhang \\
  Bureau of Geophysical Prospecting Inc.\\
  China National Petroleum Corporation\\
 Zhuozhou 072751, China \\
  \texttt{zhangtieqiang@cnpc.com.cn} \\
   \And
  Zongben Xu\\
  School of Mathematics and Statistics,\\
  Xi'an Jiaotong University\\
  Xi'an 710049, China.\\
  \texttt{zbxu@mail.xjtu.edu.cn} \\
}

\begin{document}
\maketitle
\begin{abstract}
Geophysical inverse problems are often ill-posed and admit multiple solutions. Conventional discriminative methods typically yield a single deterministic solution, which fails to model the posterior distribution, cannot generate diverse high-quality stochastic solutions, and limits uncertainty quantification. Addressing this gap, we propose an unsupervised posterior sampling method conditioned on the noisy observations and the inverse problem, eliminating the need to retrain a task-specific conditional diffusion model with paired data for each new application. Specifically, we first propose a diffusion model enhanced with a novel noise schedule for generative modeling of seismic data, and introduce the non-Markov sampling strategy to achieve fast and quality-controllable unconditional sampling. 
Building upon this, we further present a posterior sampling method for various noisy inverse problems using the trained unconditional diffusion model. Our method requires only a small number of function evaluations to achieve competitive performance, while enabling flexible posterior sampling that interacts adaptively with different noise levels.
Experiments on unconditional generation and posterior sampling across different tasks show that our method not only efficiently models the seismic data distribution and posterior conditioned on observations and tasks but also achieves substantially faster sampling and superior out-of-distribution generalization. 
\end{abstract}

% keywords can be removed
\keywords{diffusion model \and generative modeling \and posterior sampling \and accelerated sampling \and inverse problem}

\section{Introduction}
Deep learning (DL) has achieved remarkable progress in solving inverse problems across  geophysical  tasks such as denoisng \cite{yu2019deep,meng2023learning}, interpolation \cite{wang2019deep,kaur2021seismic},  imaging \cite{araya2018deep,zhang2019seismic}, reservoir parameter inversion \cite{yang2023porosity,tian2023frequency} and interpretation \cite{xiong2018seismic,wu2020building}. 
Many DL approaches follow a discriminative paradigm, producing a deterministic solution via an end-to-end neural network model. However, due to the ill-posedness of inverse problems, such deterministic solutions often fail to capture the full distribution of possible solutions and are insufficient for uncertainty quantification. From a Bayesian perspective, these solutions correspond to the minimum mean square error (MMSE) estimate \cite{kawar2021snips}, which is the conditional mean of the posterior. This averaging effect tends to oversmooth the inversion results and suppress fine structural details. In contrast, if the posterior distribution of the observation can be constructed or approximated, it becomes possible to sample multiple plausible solutions, thereby enabling a more comprehensive representation of uncertainty.

Recent advances in stochastic differential equation(SDE)-based generative models have demonstrated strong capabilities in modeling complex data distributions, supporting both unconditional and conditional generation within a unified framework for generative modeling and posterior sampling.
SDEs smoothly transforms a complex data distribution to a known prior distribution by slowly injecting noise, and a corresponding reverse-time SDE that transforms the prior distribution back into the data distribution by slowly removing the noise \cite{song2020score}. Notably, score matching with Langevin dynamics (SMLD) \cite{song2019generative} and denoising diffusion probabilistic models (DDPM) \cite{ho2020denoising} can be interpreted as special cases of exploding-variance and variance-preserving SDEs, respectively \cite{song2020score}. SMLD and DDPM have been increasingly adopted in geophysics for generative modeling of various types of seismic data (e.g., reflection data \cite{meng2024generative, meng2025generative}, velocity models \cite{wang2024controllable}, noise \cite{feng2024analysis}) and for solving seismic inverse problems, including denoising \cite{peng2024seismic,meng2025posterior,peng2025fast}, interpolation \cite{liu2024generative,wei2024seismic}, reconstruction \cite{meng2024stochastic,wang2024seisfusion,shi2024generative}, super-resolution enhancement  \cite{zhang2024seisresodiff},  coherent noise suppression  \cite{zhu2023diffusion,zhang2024conditional,durall2023deep} impedance inversion\cite{chen2025unsupervised},  imaging\cite{baldassari2024conditional}, and pretraining–finetuning paradigm for multiple tasks \cite{cheng2025agenerative}.

These studies more or less have several limitations. First, posterior sampling models are often designed for specific tasks and rely on paired labeled data for training, which limits their generalizability across different inverse problems. Second, most methods assume noise-free measurements and lack theoretical guarantees under noisy observations, making them unsuitable for real-world inversion scenarios. Consequently, models trained under fixed noise levels often fail to adapt to varying conditions. Finally, both unconditional and conditional sampling typically require hundreds to thousands of function evaluations, resulting in significant computational overhead and poor scalability for large-scale seismic applications.

To address these issues, we first proposes a diffusion model enhanced with a novel noise schedule for generative modeling of seismic data, and introduces the non-Markov sampling strategy \cite{song2020denoising} to achieve fast and quality-controllable unconditional sampling.  Then, we explore leveraging the pre-trained unconditional diffusion model for posterior sampling across various noisy inverse problems, achieving fast and efficient posterior sampling using only dozens of function evaluations.  Unlike discriminative methods, our approach generates multiple high-quality stochastic solutions, enabling users to not only select results of interest but also interactively perform posterior sampling under varying signal-to-noise ratio (SNR) conditions by adjusting the predicted noise level.
Furthermore, as the method is conditioned on both observations and forward operators, it supports flexible replacement of the forward model, making it a general and adaptable sampling-based solver for a wide range of inverse problems.

\begin{figure}[!htb]
%	\vspace{-2mm}
	\centering
	\includegraphics[width=\textwidth]{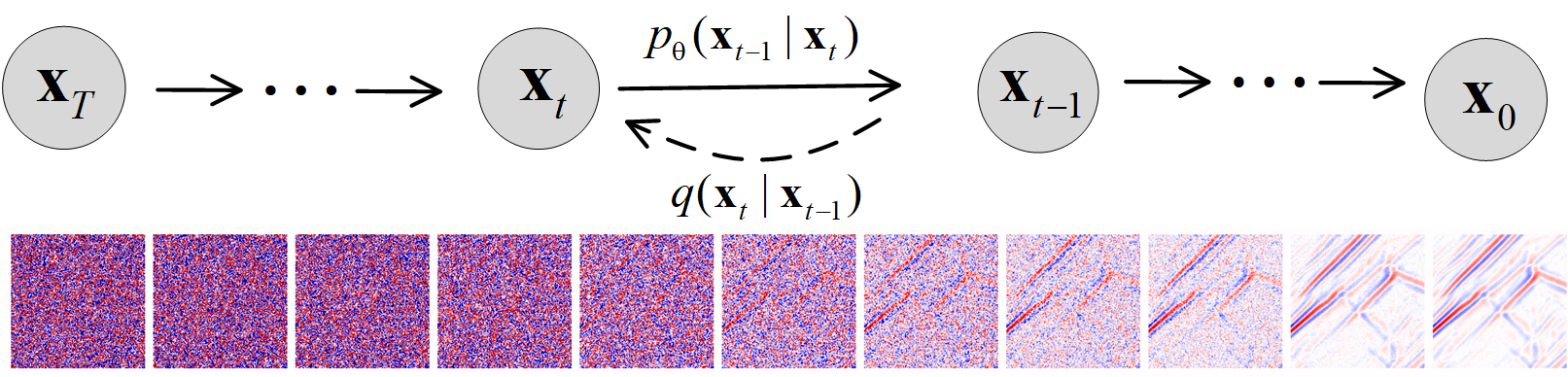}
	\caption{Schematic diagram of generative modeling for seismic data using a diffusion model.}
	\label{fig:x_evolve}
%	\vspace{-2mm}
\end{figure}

\section{Method}   

\subsection{Generative modeling of seismic data using diffusion model}
\label{subsection:optimizing}
Denoising diffusion probabilistic model (DDPM) is a class of generative model that learns to approximate complex data distributions by progressively reversing a diffusion process (see Figure \ref{fig:x_evolve}). The process consists of two main stages: a forward process (diffusion) and a reverse process (generation). 
Specifically, given data samples $\mathbf{x}_0 \sim p(\mathbf{x}_0)$, the forward diffusion process ($\mathbf{x}_0 \rightarrow \mathbf{x}_1 \rightarrow \cdots \rightarrow \mathbf{x}_T$) gradually adds noise to the data, transforming it into pure noise. This process is defined as  (Ho et al., 2020):
\begin{equation} 
	q\left(\mathbf{x}_t \mid \mathbf{x}_{t-1}\right)=\mathcal{N}\left(\mathbf{x}_t ; \sqrt{1-\beta_t} \mathbf{x}_0, \beta_t \mathbf{I}\right), t=1 \cdots T,
	\label{eq:forward-step} 
\end{equation} 
where
$0<\beta_1<\beta_2<\cdots<\beta_T<1 $ are prescribed perturbed noise variances,
$\mathcal{N}$ denotes the Gaussian distribution.
Forward process allows sampling $\mathbf{x}_{t}$ at an
arbitrary timestep $t$ in closed form:
\begin{equation} q(\mathbf{x}_t \mid \mathbf{x}_0) = \mathcal{N}(\mathbf{x}_t; \sqrt{\bar{\alpha}_t} \mathbf{x}_0, (1 - \bar{\alpha}_t) \mathbf{I}), \label{eq:marginal-distribution} 
\end{equation} 
where ${\alpha}_t=1-{\beta}_t$,$\bar{\alpha}_t = \prod_{i=1}^t \alpha_i$.
The reverse process $\mathbf{x}_T \rightarrow \mathbf{x}_{T-1} \rightarrow \cdots \rightarrow \mathbf{x}_0$ is a parameterized variational Markov chain $p_{\theta}(\mathbf{x}_{t-1} \mid \mathbf{x}_t) = \mathcal{N}(\mathbf{x}_{t-1}; \frac{1}{\sqrt{\alpha_t}} (\mathbf{x}_t - \frac{1 - \alpha_t}{\sqrt{1 - \bar{\alpha}t}} \mathbf{s}_{\theta}(\mathbf{x}_t, t)), \beta_t \mathbf{I})$. 
The training objective of $\theta$ is a variational lower bound on the negative log-likelihood of the data, which can be achieved by minimizing the following objective:
\begin{equation} 
	{L_{simple}} = 
	\mathbb{E}_{\boldsymbol{\epsilon},t,{{\bf{x}}_0}}\left[ {{{\left\| {\boldsymbol{\epsilon} - {\boldsymbol{\epsilon}_\theta }(\sqrt {{{\bar \alpha }_t}} {{\bf{x}}_0} + \sqrt {1 - {{\bar \alpha }_t}}\boldsymbol{\epsilon} ,t)} \right\|}^2}} \right]
	\label{eq:L_simple} 
\end{equation} where ${\boldsymbol{\epsilon}_\theta (\dots)}$ is the model's prediction of the noise $\boldsymbol{\epsilon} \sim \mathcal{N}({\bf{0}},{\bf{1}})$ at time step $t$.
The optimal parameter $\theta$ allows for sample generation by approximating the reverse diffusion process. The reverse process is described as follows: for $t = T, T-1, \cdots, 1$, 
\begin{equation} 
	\mathbf{x}_{t-1} = \frac{1}{\sqrt{\alpha_t}} \left( \mathbf{x}_t - \frac{1 - \alpha_t}{\sqrt{1 - \bar{\alpha}t}} \boldsymbol{\epsilon}_{\theta}(\mathbf{x}_t, t) \right) + \sigma_t \mathbf{z}, \label{eq:reverse-diffusion} 
\end{equation} 
where $\mathbf{z} \sim \mathcal{N}(\mathbf{0}, \mathbf{I})$ is an i.i.d. Gaussian noise, here $\sigma_t=\beta_t$.
\begin{figure}[htb!]
	\setlength{\abovecaptionskip}{0.3cm}
	\centering
	\subfloat[]{\includegraphics[width=0.7\textwidth]{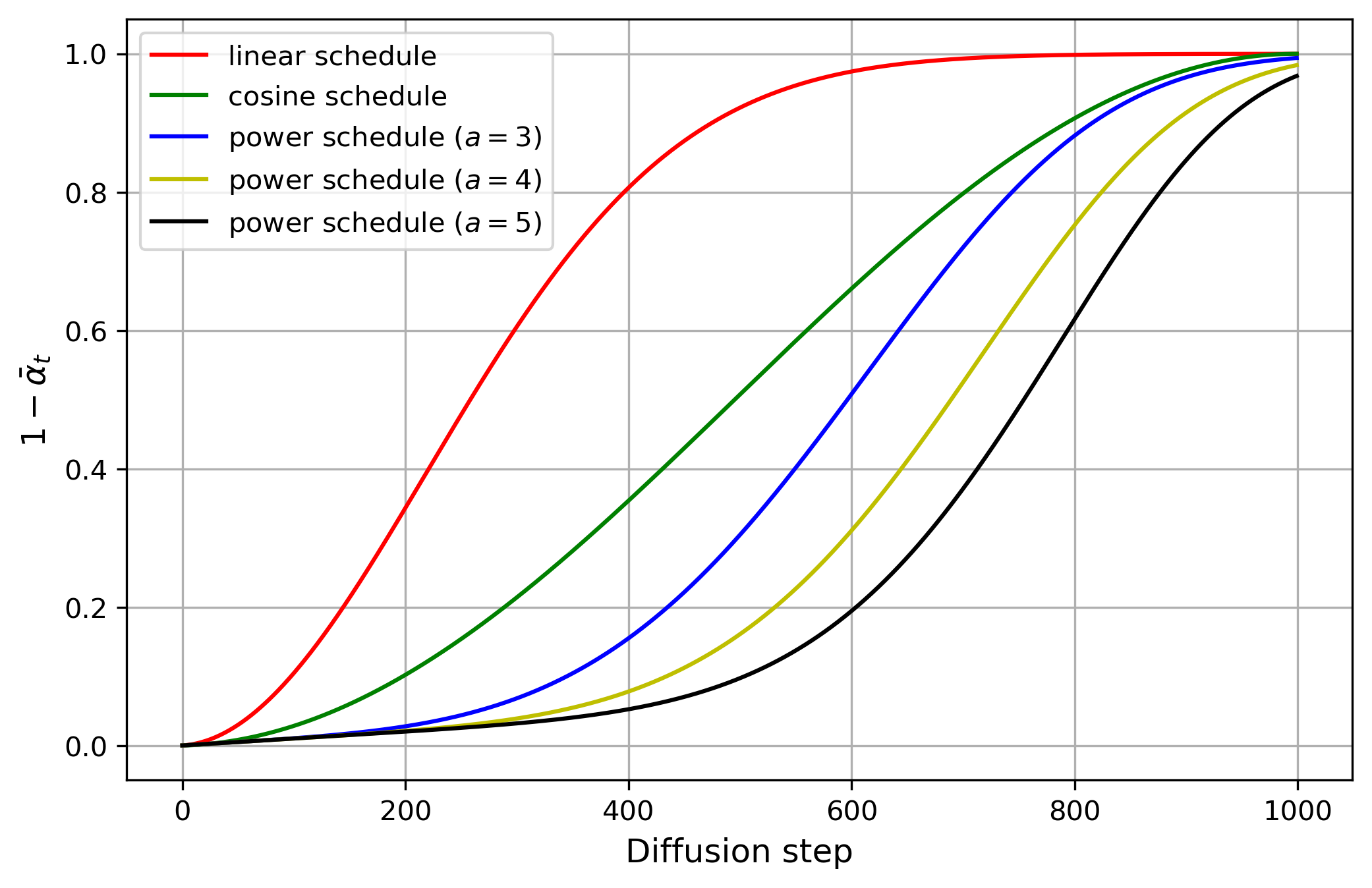}
		\label{fig:noise_schedule_curve}}\par 
	\hspace{0.4cm} 
	\subfloat[]{\includegraphics[width=\textwidth]{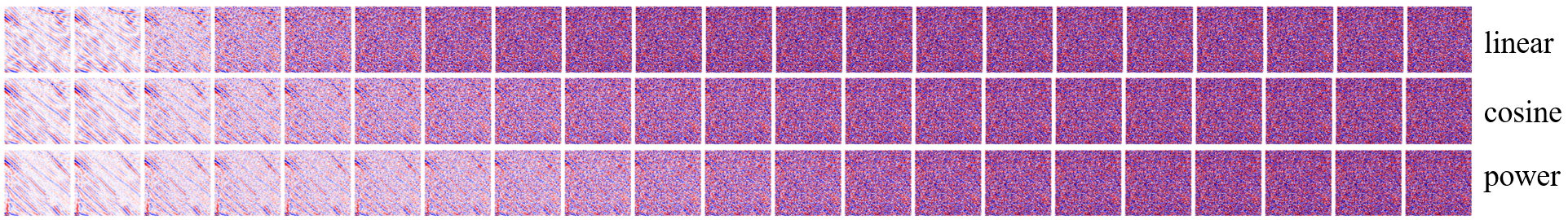}
		\label{fig:noise_schedule}}  \par        
	\caption{Comparison of linear \cite{ho2020denoising}, cosine \cite{nichol2021improved}, and power noise schedules (a=3).}
	\setlength{\belowcaptionskip}{0.3cm}
	\label{fig:noise_schedule_comp}
\end{figure}

\subsubsection{Power function noise schedule}
\begin{figure}[!htb]
%	\vspace{-1mm}
	\centering
	\includegraphics[width=\textwidth]{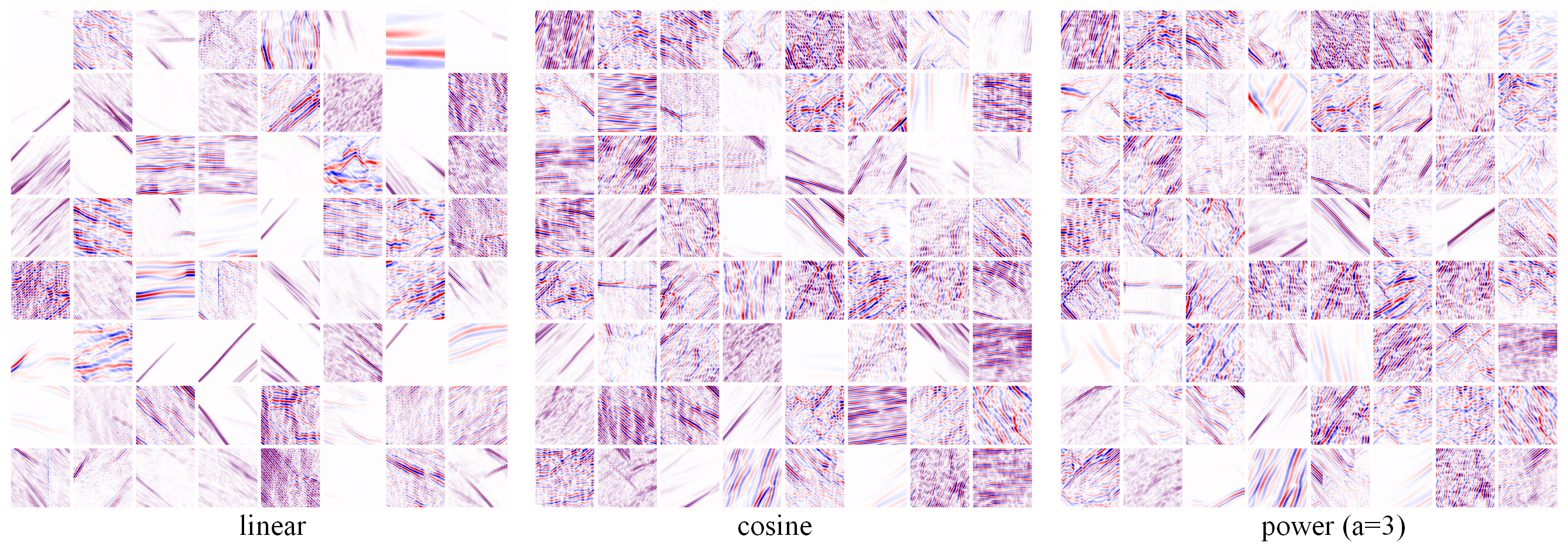}
	\caption{Results of unconditional sampling (sample generation) using a trained diffusion model with different noise strategies: linear,  cosine, and power. Each subfigure shows 64 randomly generated seismic data samples of size 128$\times$128. The sampling trajectory length is $T$ (=1000).}
	\label{fig:x_gener_nsch}
%	\vspace{-2mm}
\end{figure}

The noise schedule significantly affect the training of DDPM and the quality of samples generated by DDPM. \cite{nichol2021improved} found that injecting noise faster will result in the end of the forward noising process is too noisy, and so doesn’t contribute very much to sample quality. They then proposed a cosine noise schedule, which can inject noise more slowly than the linear noise schedule  of the original DDPM \cite{ho2020denoising}. We note that seismic signals can be buried by noise faster in the forward noising process than natural images due to their smaller mean normalized amplitude. Then we propose a new noise schedule, called the power function schedule, defined as: $\beta_t=t^a$, where $a$ is a flexible hyperparameter that enables flexible control over $\beta_t$, the larger $a$ is, the slower the noise level curve rises. The power function schedule can add noise more slowly so that the latent variables will not quickly become pure noise [see the noise level $1 - \bar{\alpha}_t$ of $\mathbf{x}_t$ in  Figure \ref{fig:noise_schedule_curve} and latent samples ($\mathbf{x}_t$) in Figure \ref{fig:noise_schedule}], which is particularly beneficial for the generative modeling of seismic signals [see Figure \ref{fig:x_gener_nsch}].

\begin{figure}[!htb]
%	\vspace{-2mm}
	\centering
	\includegraphics[width=\textwidth]{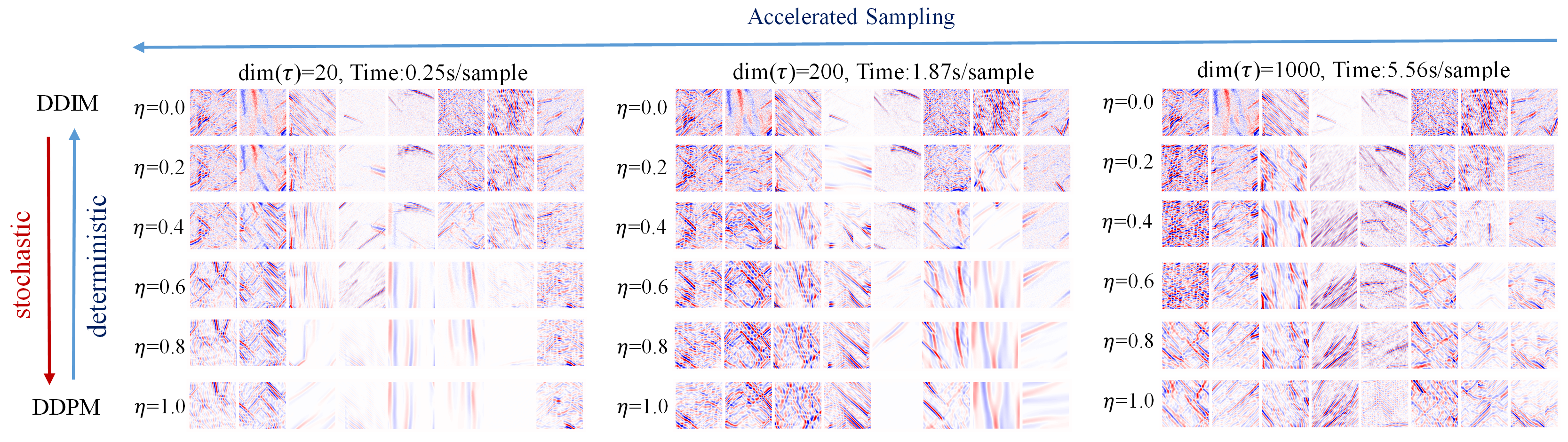}
	\caption{Unconditional sampling (sample generation) with different $\eta$ values and $dim(\tau)$. }
	
	\label{fig:eta_tau}
%	\vspace{-2mm}
\end{figure}

\subsubsection{Accelerated sampling strategy and sample quality control}
Since the sampling of a trained DDPM requires
$T$ (e.g., 1000) steps of Numbers of Function Evaluations (NFEs), which leads to long sampling times, we can also apply the principle of Denoising Diffusion Implicit Models (DDIM) \cite{song2020denoising} to accelerate the (unconditional/conditional) sampling of the trained DDPM.
DDIM is a non-Markovian sampling method that modifies the diffusion process to enable deterministic sampling, thereby reducing the number of reverse steps required. The sampling trajectory of DDIM can be represented as a subset of
$[0,1,\cdots,T]$, denoted as
$\tau$, with its length indicated by
$dim(\tau)$.
According to \cite{song2020denoising} , for $t = T, T-1, \cdots, 1$, we can generate a sample $\mathbf{x}_{t}$ from a sample $\mathbf{x}_{0}$ via
\begin{equation}
	\mathbf{x}_{t-1} = \sqrt{\bar\alpha_{t-1}} \left( \frac{\mathbf{x}_t - \sqrt{1 - \bar\alpha_t} \boldsymbol{\epsilon}_{\theta}(\mathbf{x}_t, t)}{\sqrt{\bar\alpha_t}} \right) + \sqrt{1 - \bar\alpha_{t-1}-\sigma_t^2} \boldsymbol{\epsilon}_{\theta}(\mathbf{x}_t, t)+\sigma_t\mathbf{z}
	\label{eq:ddim}
\end{equation}
where $\sigma_t=\eta\sqrt{(1-\bar\alpha_{t-1})/(1-\bar\alpha_{t})}\sqrt{(1-\bar\alpha_{t}/\bar\alpha_{t-1})}$. When $\eta=0$, formula \ref{eq:ddim} corresponds to the sampling method of DDIM. When $\eta=1$, formula \ref{eq:ddim} corresponds to the sampling method of DDPM.
Thus, the sampling speed can be effectively adjusted by choosing an appropriate sampling schedule $\tau$, as illustrated in Figure~\ref{fig:eta_tau}. When $\dim(\tau) = 20$ (NFEs = 20), the non-Markovian sampling process of DDIM can still produce high-quality and diverse results comparable to those of DDPM using the full schedule ($\dim(\tau) = 1000$, NFEs = 1000), achieving up to 50× acceleration. Furthermore, the trade-off between sample quality and diversity can be flexibly controlled by interpolating between the deterministic DDIM and the stochastic DDPM via the parameter $\eta$. As $\eta$ increases from 0 to 1, the generated samples gradually shift from being deterministic and sharp to more diverse and stochastic, reflecting a continuous trade-off between fidelity and randomness, as shown in Figure~\ref{fig:eta_tau}.

\subsection{Posterior sampling for generalized inverse problems}

In general, a supervised neural network model or conditional diffusion model are typically problem-specific and resource-intensive, requiring retraining for each task.
A promising alternative is leveraging a pre-trained DDPM (trained on domain-specific data, such as seismic data) to generate samples from the posterior distribution. For instance, a conditional model $p_\theta(\mathbf{x}_{0:T}, \mathbf{y})$ can be constructed via variational inference \cite{kawar2022denoising} or analytic conditional score function \cite{kadkhodaie2021stochastic,chung2023diffusion, meng2024stochastic}, reusing the same learnable parameters ($\theta$) from the pre-trained DDPM $p_\theta(\mathbf{x}_{0:T})$, thus avoiding the need for retraining.

\subsubsection{Multi-purpose posterior sampling using a trained diffusion model}

\begin{figure}[!htb]
	%	\vspace{-2mm}
	\centering
	\includegraphics[width=\textwidth]{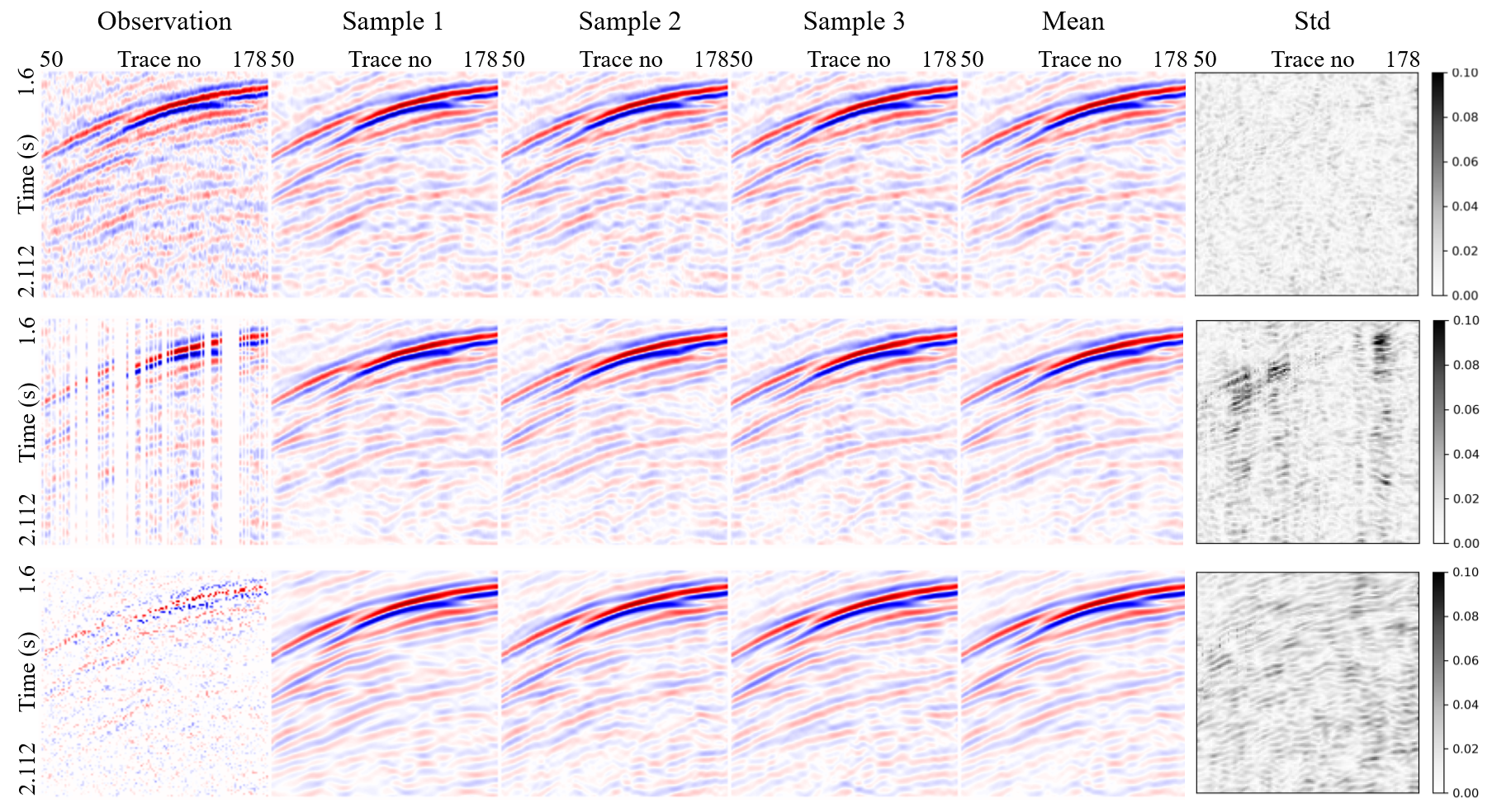}
	\caption{An example of multi-task posterior sampling results, taking denoising, interpolation, and compressed sensing as examples. `Mean' represents the posterior expectation (the mean of the sample), and std represents the standard deviation of the samples. }
	\label{fig:task_field}
	%	\vspace{-2mm}
\end{figure}

Given an observation $\mathbf{y}$, its posterior distribution can be expressed as $p(\mathbf{x}|\mathbf{y})$, where  
\begin{equation} 
	\mathbf{y} = \mathbf{G}\mathbf{x} + \mathbf{n}, \label{eq:y} 
\end{equation}
$\mathbf{G}$ is the forward operator, and $\mathbf{n}$ represents noise.
To connect the noise in the observation $\mathbf{y}$ with the diffusion noise in $\mathbf{x}_{1:T}$, we perform diffusion in the spectral domain of $\mathbf{G}$ via its singular value decomposition (SVD) inspired by \cite{kawar2022denoising}. For a general linear $\mathbf{G}$, its SVD is given as 
\begin{align}
	\mathbf{G} = \mathbf{U} \mathbf{\Sigma} \mathbf{V}^\top, 
\end{align}
where $\mathbf{U} \in \mathbb{R}^{m \times m}$, $\mathbf{V} \in \mathbb{R}^{n \times n}$ are orthogonal matrices, and $\mathbf{\Sigma} \in \mathbb{R}^{m \times n}$ is a rectangular diagonal matrix containing the singular values of $\mathbf{G}$, ordered descendingly. 
We denote the singular values as $s_1 \geq s_2 \geq \ldots \geq s_{m}$, and define $s_{i} = 0$ for $i \in [m+1, n]$.
Then, we employ a new Markov model conditioned on the observation $p_\theta({\mathbf{x}}_{0:T} \mid \mathbf{y})$ following \cite{kawar2022denoising}, which leverages only the learnable parameters $\theta$ of a pre-trained unconditional diffusion model to perform posterior sampling without requiring any additional training.  For ease of derivation and notation, the variational distribution $q$ of the model is constructed such that $q(\mathbf{x}_t \mid \mathbf{x}_0) = \mathcal{N}(\mathbf{x}_0, \hat{\sigma}_t^2 \boldsymbol{I})$, where the noise levels satisfy $0 = \hat{\sigma}_0 < \hat{\sigma}_1 < \hat{\sigma}_2 < \ldots < \hat{\sigma}_T$.  $\hat{\sigma}_t$ denotes the diffusion noise level of ${\mathbf{x}}_t$ relative to $\mathbf{x}_0$, which differs from the predefined $\sigma_t$ in Formula \ref{eq:reverse-diffusion}.
The variables $\hat{\sigma}_t$  satisfy the relations $\hat{\sigma}_t = \sqrt{(1 - \alpha_t)/\alpha_t}$, the derivation of which can be found in \cite{song2020score,kawar2022denoising}.
Here, the new variable ${\mathbf{x}}_t$ is redefined as ${\mathbf{x}}_t=\hat{\mathbf{x}}_t \sqrt{1 + \sigma_t^2}= \hat{\mathbf{x}}_t/\sqrt{\alpha_t}$, where $\hat{\mathbf{x}}_t$ corresponds to ${\mathbf{x}_t}$ from the forward process in DDPM (see formula~\ref{eq:forward-step}).
Then, given a noisy observation $\mathbf{y}$ with noise level $\sigma_{\mathbf{y}}$,
the reverse-time diffusion process of the diffusion model conditioned on $\mathbf{y}$, denoted as $p_\theta^{(t)}({\mathbf{x}}_t \mid {\mathbf{x}}_{t+1}, \mathbf{y})$, can be formulated as follows.
When $t = T$, the process starts by initializing $\hat{\mathbf{x}}_T$ as:
\begin{align}
	\bar{\mathbf{x}}_T^{(i)} =
	\begin{cases}
		\bar{\mathbf{y}}_i + \left( \hat{\sigma}_{T}^2 - \frac{\sigma_{\mathbf{y}}^2}{s_i^2} \right) \mathbf{z} & \text{if } s_i > 0, \\
		\hat{\sigma}_{T}^2 \mathbf{z} & \text{if } s_i = 0,
	\end{cases}
	\label{eq:pt-init}
\end{align}
where $\bar{\mathbf{y}}^{(i)}$ denotes the $i$-th entry of $\bar{\mathbf{y}} = \mathbf{\Sigma}^\dagger \mathbf{U}^\top \mathbf{y}$.
The subsequent steps proceed as follows: for $t = T-1, \ldots, 1$, the transition is given by:
\begin{align}
	\bar{\mathbf{x}}_t^{(i)} =
	\begin{cases}
		\bar{\mathbf{x}}_{0,t}^{(i)} + \sqrt{1 - \gamma^2} \hat{\sigma}_{t} \frac{\bar{\mathbf{x}}_{t+1}^{(i)} - \bar{\mathbf{x}}_{0,t}^{(i)}}{\sigma_{t+1}}+\gamma^2 \hat{\sigma}_{t}^2\mathbf{z} & \text{if } s_i = 0, \\
		\bar{\mathbf{x}}_{0,t}^{(i)} + \sqrt{1 - \gamma^2} \hat{\sigma}_{t} \frac{\bar{\mathbf{y}}_i - \bar{\mathbf{x}}_{0,t}^{(i)}}{\sigma_{\mathbf{y}} / s_i}+ \gamma^2 \hat{\sigma}_{t}^2\mathbf{z} & \text{if } \sigma_{t} < \frac{\sigma_{\mathbf{y}}}{s_i}, \\
		(1 - \gamma_b) \bar{\mathbf{x}}_{0,t}^{(i)} + \gamma_b \bar{\mathbf{y}}_i+ (\hat{\sigma}_{t}^2 - \frac{\sigma_{\mathbf{y}}^2}{s_i^2} \gamma_b^2)\mathbf{z} & \text{if } \hat{\sigma}_{t} \geq \frac{\sigma_{\mathbf{y}}}{s_i}.
	\end{cases}
	\label{eq:pt-transition}
\end{align}
where $\mathbf{x}_{0,t}=\left(\hat{\mathbf{x}}_{t+1}-\sqrt{1-\bar{\alpha}_{t+1} }\boldsymbol{\epsilon}_{\theta}(\hat{\mathbf{x}}_{t+1}, t)\right)/\sqrt{\bar{\alpha}_t}$ are the predictions of $\mathbf{x}_0$ at every step $t$ using the trained unconditional diffusion model (see Formula \ref{eq:marginal-distribution}). ${\bar{\mathbf{x}}_{0, t}^{(i)}}$ and ${\bar{\mathbf{x}_t}^{(i)}}$ denote the $i$-th entries of ${\bar{\mathbf{x}}_{0, t}} = \mathbf{V}^\top \mathbf{x}_{0, t}$ and ${\bar{\mathbf{x}}_t} = \mathbf{V}^\top \mathbf{x}_t$,  respectively.  $\dagger$ denotes the Moore-Penrose pseudo-inverse. $\sigma_{\mathbf{y}}$ is the noise level of the observation, which is an unknown scalar to be estimated. The variables with a bar represent vectors in the spectral domain, and can be transformed back into  vectors in the pixel space by left-multiplying with the matrix $\mathbf{V}$, as $\mathbf{x}_t=\mathbf{V}\bar{\mathbf{x}}_t$.
$\gamma \in (0, 1]$ and  $\gamma_b \in (0, 1]$ are two hyperparameters, which control the level of noise injected at each time step in reverse diffusion processes. We tested the effects of different hyperparameters values ($\gamma$ and $\gamma_b $) on the sampling results, which were relatively stable and insensitive. Therefore, we kept these two hyperparameters unchanged in the experiments throughout the paper, setting them to $\gamma$=0.9 and $\gamma_b$=1.  The full sampling procedure is provided in Algorithm \ref{Alg:PS}. In Figure \ref{fig:task_field}, we present an example of posterior sampling for different tasks, showing three stochastic solutions along with their mean and standard deviation, which are useful for solution selection and uncertainty quantification.
\begin{algorithm}
	\caption{Generalized posterior sampling for noisy inverse problems using the trained unconditional DM}
	\begin{algorithmic}[1]
		\renewcommand{\algorithmicrequire}{\textbf{Input:}}
		\renewcommand{\algorithmicensure}{\textbf{Output:}}
		\REQUIRE $\boldsymbol{\epsilon}_{\theta}$, $\sigma_{\mathbf{y}}$, $\mathbf{y}$, $\mathbf{G}$, $\gamma$, $\gamma_b$. 
		\ENSURE  ${\mathbf{x}}_0$;
		\STATE  Initialize $\bar{\bf{x}}_T \gets \mathbf{V}^\top \mathbf{x}_T$ \COMMENT {using Formula \ref{eq:pt-init}}.
		\STATE  $\mathbf{U}, {\Sigma}, \mathbf{V}^\top  \gets  \mathbf{G} = \mathbf{U} \mathbf{\Sigma} \mathbf{V}^\top $.
		\FOR {$t  \gets T-1$ to $1$} 
		\STATE{$\hat{\mathbf{x}}_{t+1} \gets \sqrt{\alpha_{t+1}}{\mathbf{x}}_{t+1} $}.
		\STATE {$\mathbf{x}_{0,t} \gets \left(\hat{\mathbf{x}}_{t+1}-\sqrt{1-\bar{\alpha}_{t+1} }\boldsymbol{\epsilon}_{\theta}(\hat{\mathbf{x}}_{t+1}, t)\right)/\sqrt{\bar{\alpha}_{t+1}}$ }.
		\STATE {${\bar{\mathbf{x}}_{0,t}} \gets   \mathbf{V}^\top \mathbf{x}_{0,t}$}.
		\STATE{${\mathbf{x}}_{t+1} \gets \hat{\mathbf{x}}_{t+1}/\sqrt{\alpha_{t+1}} $}
		\STATE {${\bar{\mathbf{x}}_{t+1}} \gets   \mathbf{V}^\top \mathbf{x}_{t+1}$}. 	\COMMENT {transform to spectral space}
		\STATE Draw ${\bf{z}} \sim \mathcal{N}(\bm{0}, \bm{I})$.
		\STATE Obtain $\bar{\mathbf{x}}_t$ using Formula \ref{eq:pt-transition}.
		\STATE {${\mathbf{x}}_t \gets   \mathbf{V} \bar{\mathbf{x}}_t$}. 	\COMMENT {transform to pixel space}
		\ENDFOR
		\RETURN ${\mathbf{x}}_0$.
	\end{algorithmic}
	\label{Alg:PS} 
\end{algorithm}

\subsubsection{Accelerated posterior sampling}
Note that an optimal solution to DDPM or DDIM is also optimal for the conditional diffusion model $p_\theta({\mathbf{x}}_{0:T} \mid \mathbf{y})$ discussed earlier, since they share the same learnable parameter $\theta$, as shown in \cite{kawar2022denoising}. Consequently, the accelerated sampling strategies developed for DDPM and DDIM are equally applicable in this conditional setting.

\begin{figure}[!htb]
%	\vspace{-2mm}
	\centering
	\includegraphics[width=\textwidth]{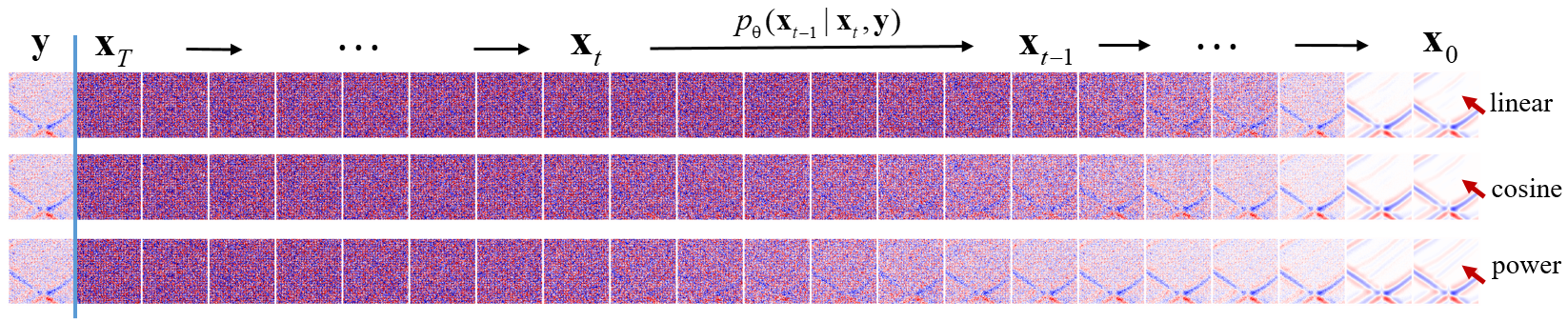}
	\caption{ The impact of diffusion models trained with different noise schedules on posterior sampling results. }
	\label{fig:ps_ns_comp}
%	\vspace{-2mm}
\end{figure}

%the forward diffusion process of the conditional diffusion model is defined as
%$q\left(\mathbf{x}_t \mid \mathbf{x}_0\right) = \mathcal{N}\left(\mathbf{x}_0, \sigma_t^2 \boldsymbol{I} \right)$ for $t = 1,  \ldots, T-1, T$,
%where $\mathbf{x}_t = \hat{\mathbf{x}}_0 + \hat{\sigma}_t^2 \mathbf{z}$  and $\mathbf{z} \sim \mathcal{N}(\mathbf{0}, \mathbf{I})$.
%Here, $\hat{\sigma}_t$ denotes the diffusion noise level of $\hat{\mathbf{x}}_t$ relative to $\mathbf{x}_0$, which differs from the predefined $\sigma_t$ in Formula \ref{eq:reverse-diffusion}.
%The variable $\hat{\sigma}_t$ satisfies the relation $\sigma_t = \sqrt{(1 - \alpha_t) / \alpha_t}$, the derivation of which can be found in \cite{song2020score,kawar2022denoising}.

\subsubsection{Enhanced posterior sampling using DDPM trained with power noise schedule}
Different noise schedules will affect the diversity and quality of unconditional sampling samples [see Figure \ref{fig:x_gener_nsch}], and will also affect the quality of the posterior sampling solution. As shown in Figure \ref{fig:ps_ns_comp}, the learnable parameter $\theta$ obtained by training with the proposed power function noise schedule will enable the posterior sampling model ($p_\theta(\mathbf{x}_t | \mathbf{x}_{t+1}, \mathbf{y})$) to obtain a higher quality posterior solution

\begin{figure}[!htb]
%	\vspace{-2mm}
	\centering
	\includegraphics[width=\textwidth]{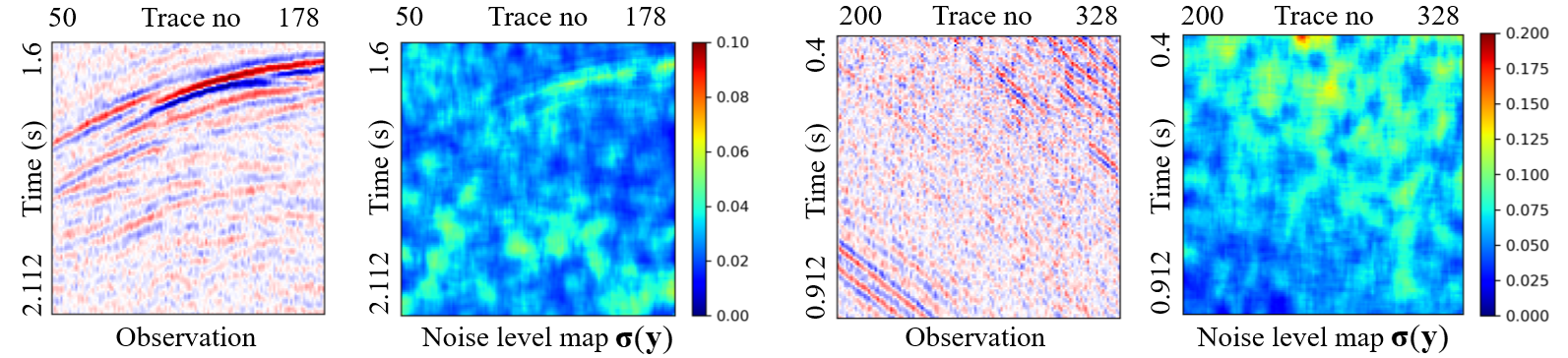}
	\caption{Field data and corresponding pixelwise noise level map.}
	\label{fig:noise level map}
%	\vspace{-2mm}
\end{figure}

\subsubsection{Interaction with noise level}
For noisy seismic inverse problems, accurately estimating the noise level $\sigma_\mathbf{y}$ in $\mathbf{y}$ is crucial for effectively generating samples from the posterior using diffusion models. To address this, we employ a variational inference model (called as Variational Inference non-independent and non-identically distributed (VI-non-IID)) trained on seismic data \cite{meng2022seismic} to directly predict $\sigma_\mathbf{y}$.
Let $\boldsymbol{\sigma}_\mathbf{y}$ represent a pixelwise noise level with the same shape as $\mathbf{y}$ [see Figure \ref{fig:noise level map} for an example of pixelwise noise levels on feild data].
For stationary noise, such as IID Gaussian noise, 
$\sigma_\mathbf{y}$ can be set as 
$median(\boldsymbol{\sigma}_\mathbf{y})$, for non-stationary noise, $\sigma_\mathbf{y}$ can be interactively selected within the interval $[min(\boldsymbol{\sigma}_\mathbf{y}),max(\boldsymbol{\sigma}_\mathbf{y})]$.

\begin{figure}[!htb]
%	\vspace{-2mm}
	\centering
	\includegraphics[width=\textwidth]{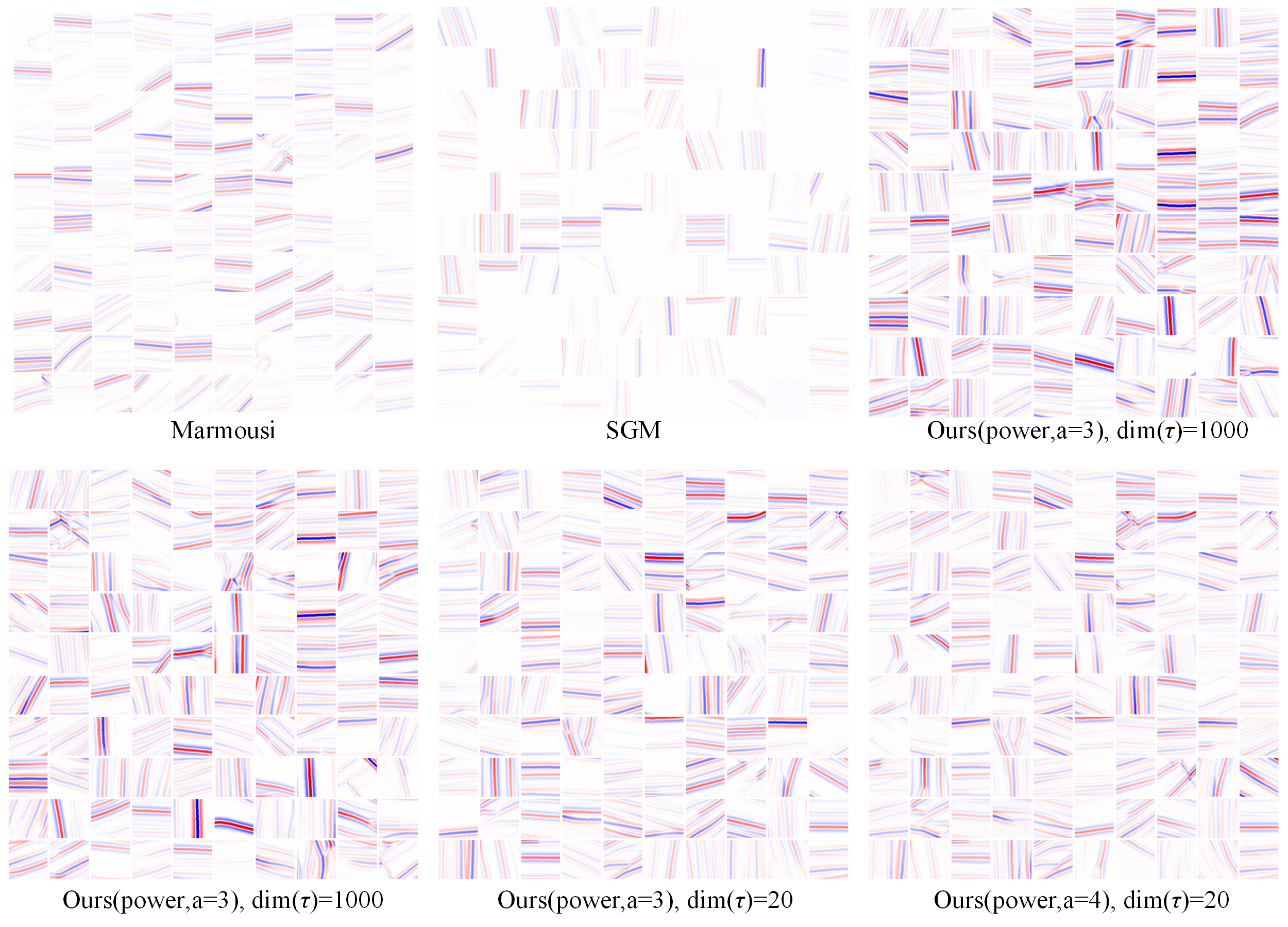}
	\caption{Unconditional sampling results of different generative models. Marmousi represents samples randomly sampled from the real marmousi dataset.}
	\label{fig:gener_com}
%	\vspace{-2mm}
\end{figure}

\section{Examples}

%\subsection{Experimental setup}

\subsection{Unconditional sampling}
We take the Marmousi dataset as an example to demonstrate the effectiveness of our method in seismic data generative modeling. As shown in Figure~\ref{fig:gener_com}, we present 100 randomly sampled Marmousi images of size $128 \times 128$, which simulate samples drawn from the ground-truth data distribution $p(\mathbf{x})$. In addition, we provide unconditional samples generated by various generative models to approximate the learned data distribution $p(\mathbf{x})$.
Specifically, SGM refers to the score-based generative model proposed in~\cite{song2019generative,song2020improved}, which can be interpreted as a variance-exploding stochastic differential equation (SDE) driven by a neural network \cite{song2020score}. We adopt the open-source SGM model trained on seismic data provided by~\cite{meng2024generative} to generate 100 samples. The number of function evaluations (NFEs) is 500, and the total sampling time is 3951 seconds.
Moreover, we provide sampling results from our proposed method (here $\eta$=0.8), which is based on a DDPM trained with a power noise schedule. The parameter $a$ controls the shape of the noise schedule, and we present results for $a = 3$ and $a = 4$. The diffusion step size is controlled by $dim(\tau)$, where $dim(\tau)$ = 1000 corresponds to the full diffusion trajectory, and $dim(\tau)$ = 20 corresponds to a sub-sequence of the full trajectory. The total sampling times for $dim(\tau)$ = 1000 and $dim(\tau)$ = 20 are 2393 seconds and 39 seconds, respectively. As observed in Figure~\ref{fig:gener_com}, our method is capable of generating samples with greater diversity, including those with complex geological structures. Remarkably, the accelerated sampling strategy, which achieves over a 50$\times$ speed-up, still produces samples of comparable quality to those generated with the full sampling trajectory.

To capture the distribution of a broader range of seismic datasets, we further trained new DDPM models using the Marmousi and the publicly available OpenSEGY\footnote{https://wiki.seg.org/wiki/Open\_data\#2D\_synthetic\_seismic\_data} synthetic datasets, which have also been used to train other generative and denoising models\cite{meng2024generative,meng2022seismic,yu2019deep}. The sampling results are shown in Figure~\ref{fig:x_gener_nsch} and Figure~\ref{fig:eta_tau}.
In the following experiments, all employed generative models and open-source comparative generative methods (score-based generative models, SGM\cite{meng2024generative}) are trained on this dataset.  All experiments were conducted on a single NVIDIA RTX 3090 GPU with 24 GB of memory. We trained the unconditional DDPM for a total of 30,000 iterations, with a total training time of 26.5 hours.

\begin{figure}[!htb]
%	\vspace{-2mm}
	\centering
	\includegraphics[width=\textwidth]{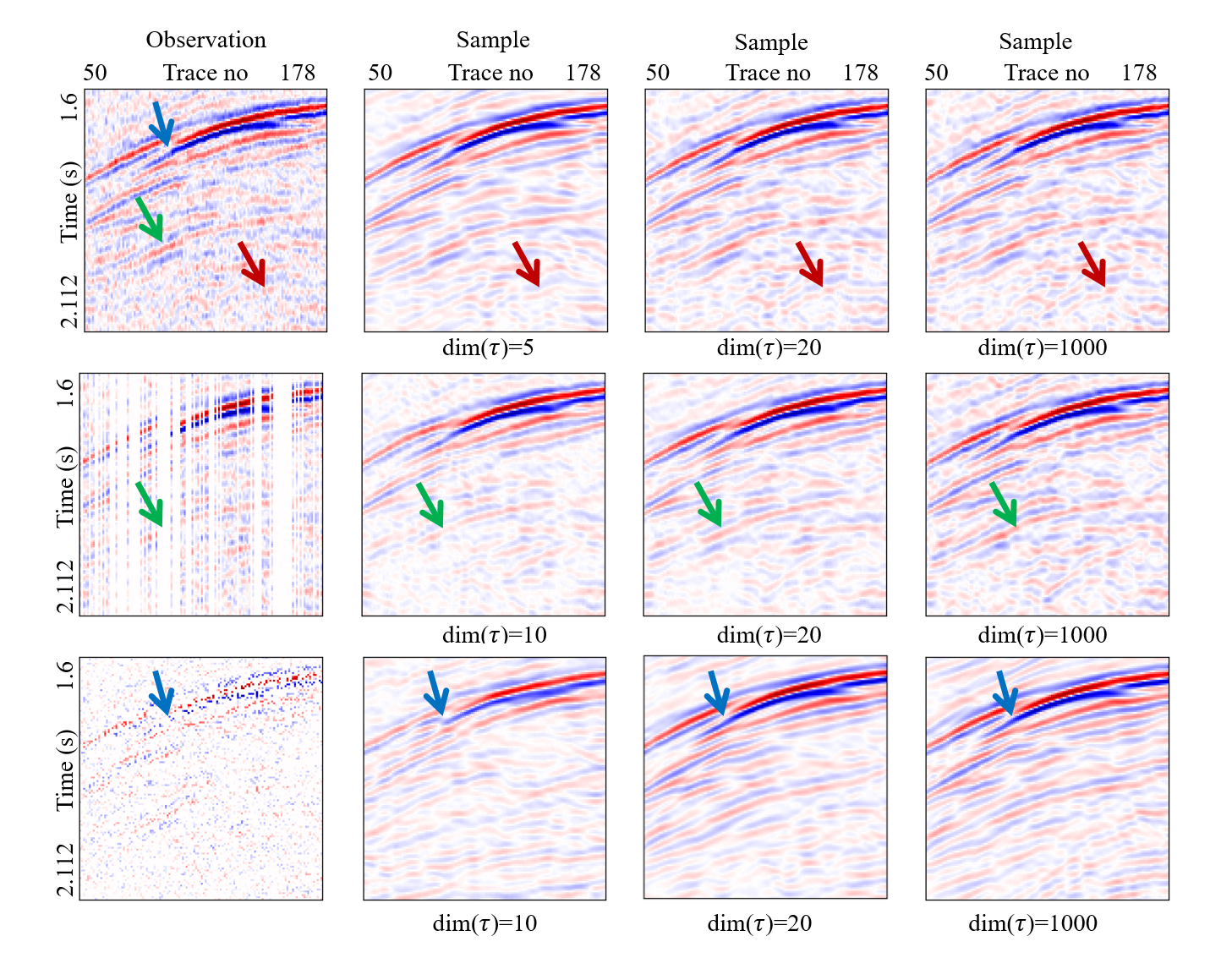}
	\caption{Posterior sampling results of subsequence trajectories of different lengths}
	\label{fig:dim_tau}
%	\vspace{-2mm}
\end{figure}

\subsection{Accelerated posterior sampling}
We consider three representative inverse problems under varying noise conditions: denoising, interpolation, and compressed sensing, to evaluate the effectiveness of our accelerated posterior sampling approach using subsequences of different lengths.
For the denoising task, we set $\mathbf{G} = \mathbf{I}$, where $\mathbf{I}$ is the identity matrix.  
For the interpolation task, $\mathbf{G} = \mathbf{M}$, where $\mathbf{M}$ is a binary mask matrix indicating the positions of missing entries.  
For the compressed sensing task, $\mathbf{G} = \mathbf{P} \mathbf{W}$, where $\mathbf{P} \in \mathbb{R}^{d_0 \times d}$ is a random sampling operator and $\mathbf{W} \in \mathbb{R}^{d \times d}$ is a transformation matrix. Here, we use Fast Walsh–Hadamard Transform\cite{fino1976unified} to implementate $\mathbf{W}$ .The observation model is expressed as
%$\mathbf{y} = \mathbf{P} \mathbf{W} \mathbf{x} + \mathbf{n}$,
\begin{equation} 
	\mathbf{y} =\mathbf{G} \mathbf{x} + \mathbf{n}= \mathbf{P} \mathbf{W} \mathbf{x} + \mathbf{n}, \label{eq:y_cs} 
\end{equation}
where $\mathbf{y} \in \mathbb{R}^{d_0}$ and $d_0 < d$, with $\mathbf{n}$ denoting additive noise, ${d_0}/d$ is is the compression ratio.
For better visualization, in Figure~\ref{fig:dim_tau}, the measurements $\mathbf{y}$ for the compressed sensing task are illustrated by assigning the observed (randomly sampled via $\mathbf{P}$) entries to their respective spatial locations, while unobserved entries are filled with zeros.
The field data used for evaluation is shown in Figure~\ref{fig:noise level map}. In the interpolation task, 50\% of seismic traces are randomly missing. In the compressed sensing task, the measurements are acquired with a compression ratio of 25\%.

As shown in Figure~\ref{fig:dim_tau}, our method achieves results comparable to those obtained using 1000 sampling steps, while requiring only a few dozen steps, demonstrating substantial acceleration without sacrificing sampling quality.

\begin{figure}[htb!]
	\setlength{\abovecaptionskip}{0.3cm}
	\centering
	% 第一行 (a)
	\subfloat[]{\includegraphics[width=0.9\textwidth]{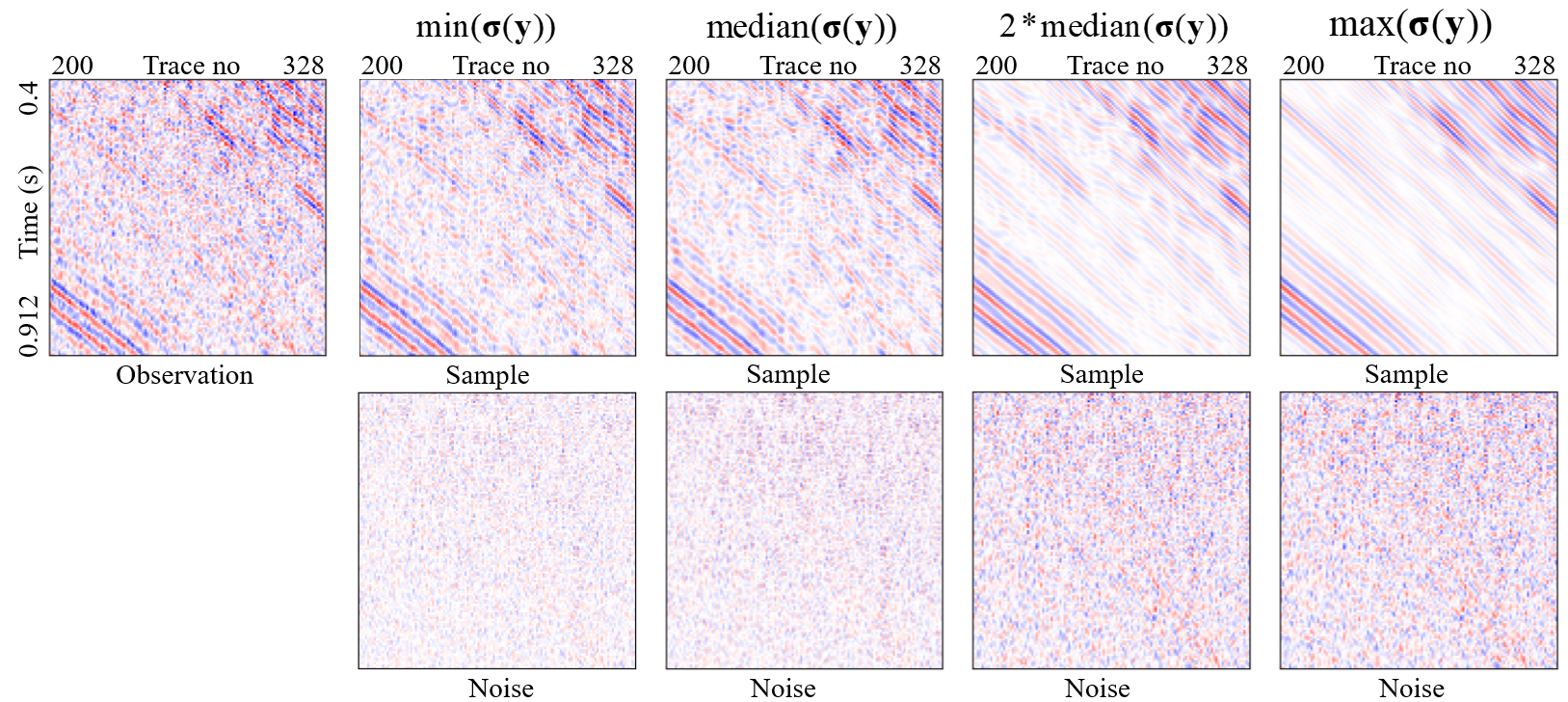}
		\label{fig:noise_level_inter_dn}}\\[0.4cm]
	% 第二行 (b)
	\subfloat[]{\includegraphics[width=0.7\textwidth]{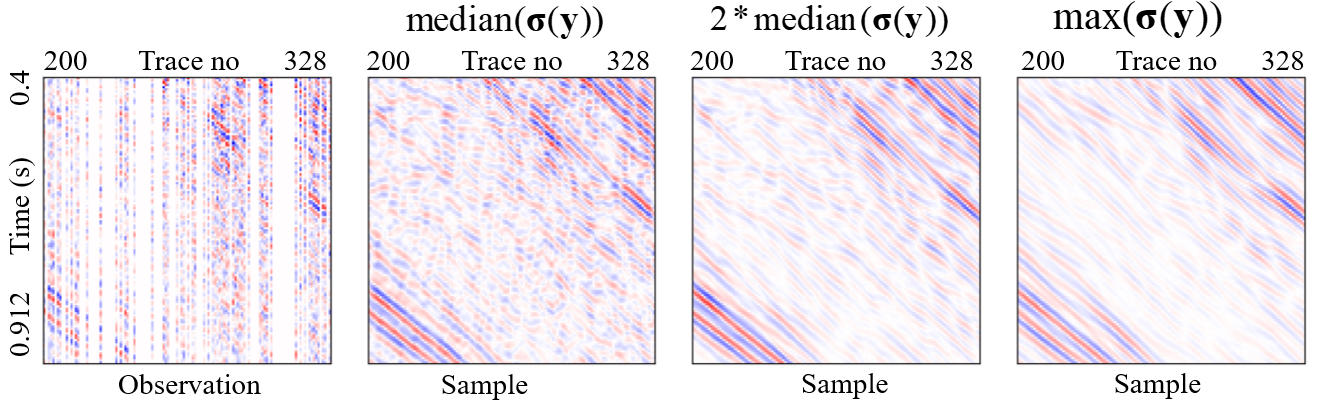}
		\label{fig:noise_level_inter_int}}\\[0.4cm]
	% 第三行 (c)
	\subfloat[]{\includegraphics[width=0.7\textwidth]{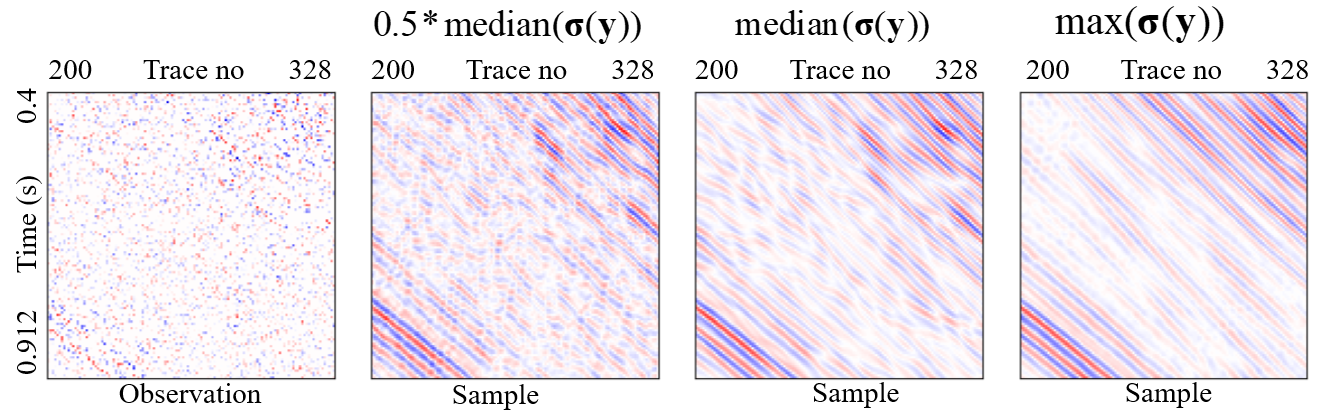}
		\label{fig:noise_level_inter_cs}}
	\caption{Posterior sampling interacting with predicted noise level. Take (a) denoising, (b) interpolation, and (c) compressed sensing as examples.}
	\label{fig:noise_level_inter}
\end{figure}

\subsection{Posterior sampling interacting with noise level}
In our posterior sampling method, the noise level in the observation is a hyperparameter that must be estimated. Setting different input noise levels not only enables the posterior sampling to produce high-quality inversion results but also controls the strength of noise suppression. This choice is not a burden; rather, it provides flexibility and supports interactive posterior sampling adaptable to practical scenarios. Figure ~\ref{fig:noise_level_inter} presents the posterior sampling results under different noise level settings, using the same three inverse problems as examples: denoising, interpolation, and compressed sensing.
The field data, shown in Figure~\ref{fig:noise level map}, is severely corrupted by noise that is non-independent and identically distributed (non-i.i.d.) and exhibits spatially varying noise levels. The pixelwise noise levels predicted by VI-non-IID show that the minimum, median, and maximum values of $\boldsymbol{\sigma}(\mathbf{y})$, denoted as $min(\boldsymbol{\sigma}(\mathbf{y}))$, $median(\boldsymbol{\sigma}(\mathbf{y}))$, and $max(\boldsymbol{\sigma}(\mathbf{y}))$, are 0.0204, 0.0658, and 0.1739, respectively.
Taking the denoising task as an example, as shown in Figure~\ref{fig:noise_level_inter_dn}, when $\sigma(\mathbf{y}) = \min(\boldsymbol{\sigma}(\mathbf{y}))$, the noise suppression is weakest. As $\sigma(\mathbf{y})$ increases, the suppression strength also increases. Using the default value $\sigma(\mathbf{y}) = \mathrm{median}(\boldsymbol{\sigma}(\mathbf{y}))$ yields moderate suppression, while $\sigma(\mathbf{y}) = \max(\boldsymbol{\sigma}(\mathbf{y}))$ results in the strongest suppression. In all cases, the denoised sections exhibit minimal or no visible signal leakage, indicating that the energy of the useful signals is well preserved. Users can not only interactively sample different posterior solutions under a fixed $\sigma(\mathbf{y})$, but also adjust $\sigma(\mathbf{y})$ to explore varying degrees of noise suppression, thereby enabling a flexible and user-controllable posterior sampling process.
The same phenomenon can also be observed in the interpolation and compressed sensing tasks, as shown in Figure~\ref{fig:noise_level_inter_int} and ~\ref{fig:noise_level_inter_cs}.

\begin{figure}[!htbp]
	\setlength{\abovecaptionskip}{0.0cm}
	\centering
	% 第一张图 (a)
	\subfloat[]{\includegraphics[width=\textwidth]{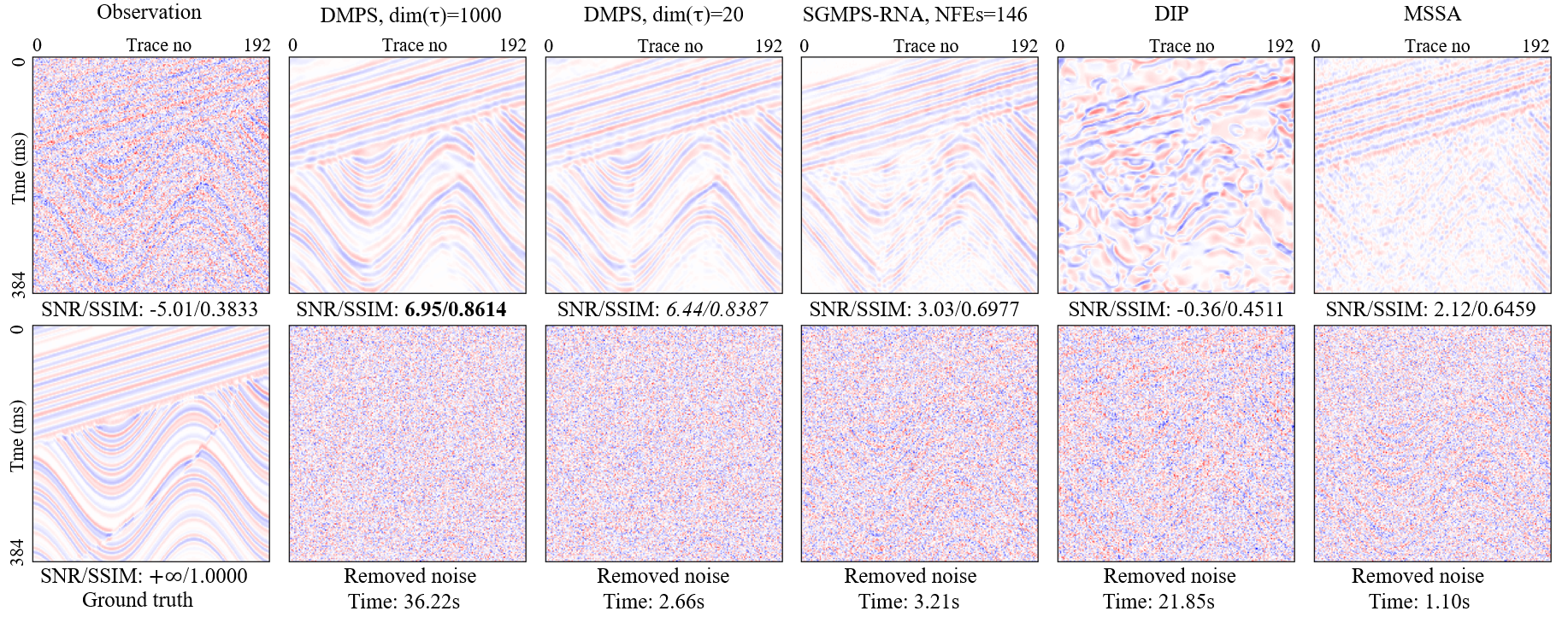}
		\label{fig:sigmoid_dn}}\\[0.0cm]
	% 第二张图 (b)
	\subfloat[]{\includegraphics[width=\textwidth]{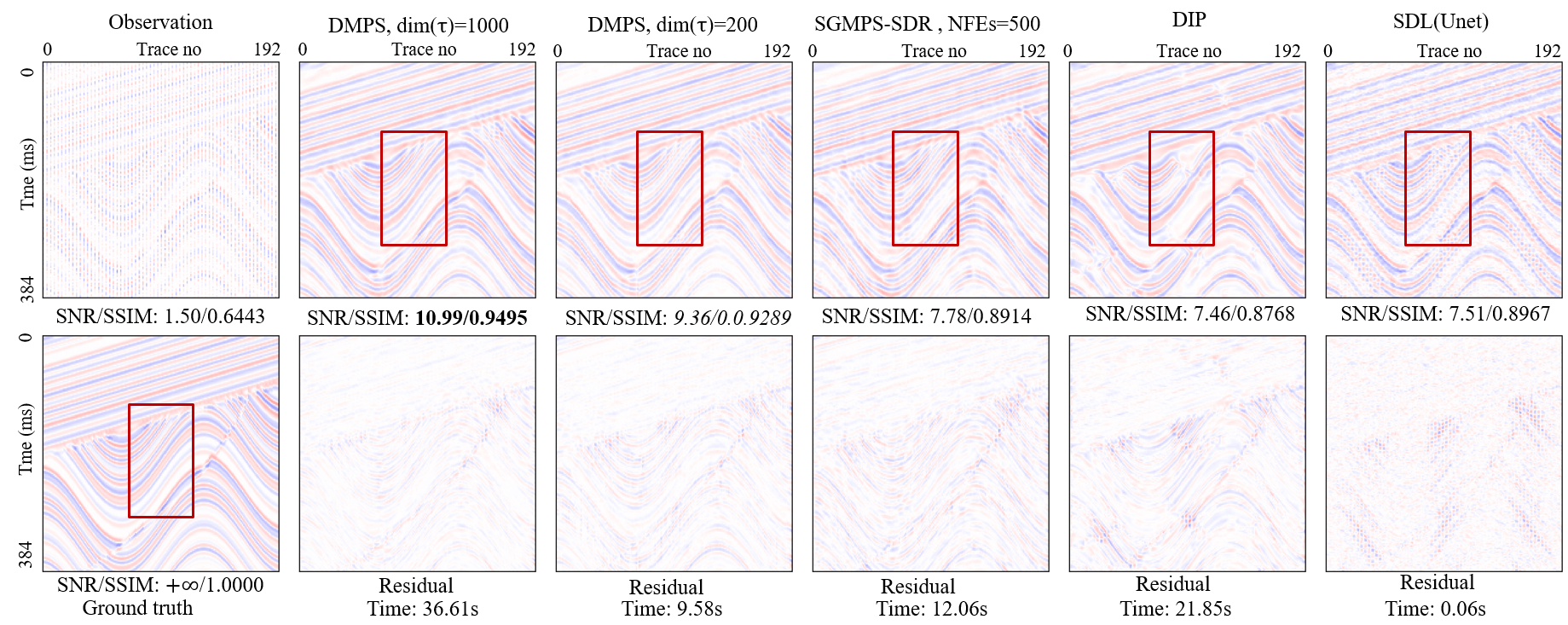}
		\label{fig:sigmoid_int}}\\[0.0cm]
	% 第三张图 (c)
	\subfloat[]{\includegraphics[width=\textwidth]{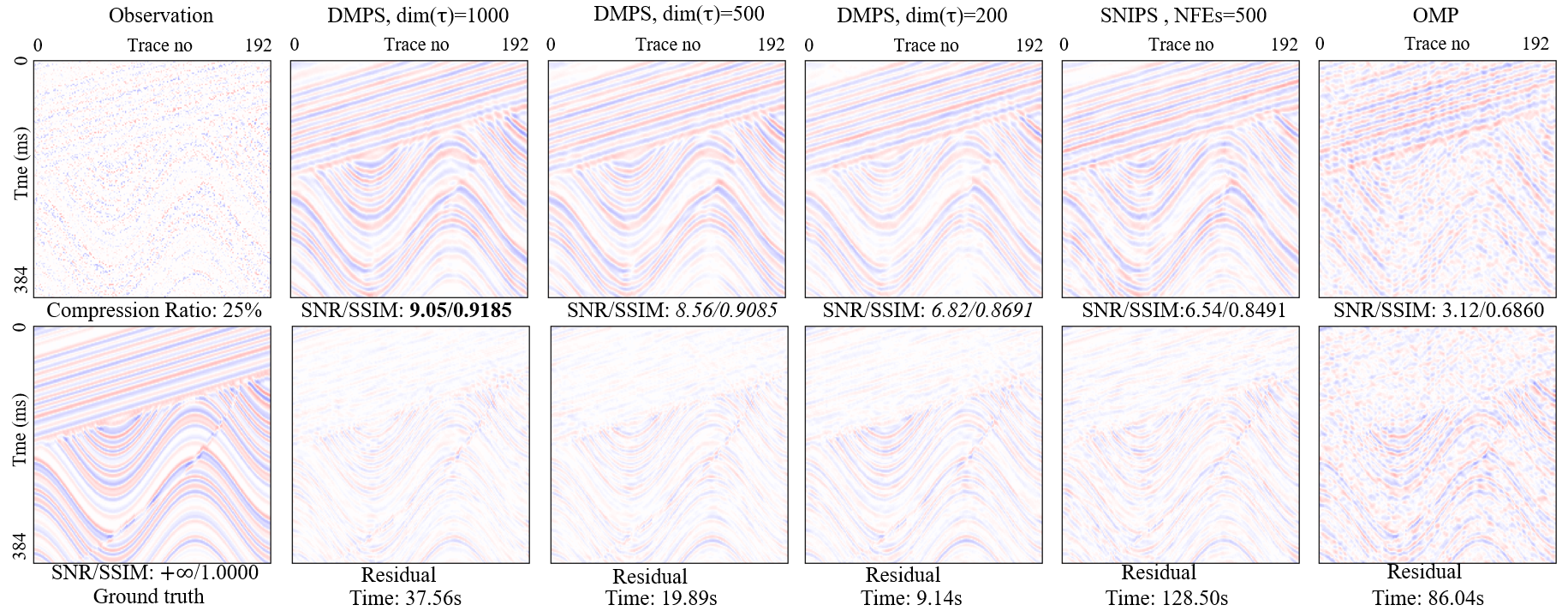}
		\label{fig:sigmoid_cs}}
	\caption{Posterior sampling results of synthetic data, taking tasks (a) denoising, (b) interpolation, and (c) compressed sensing as an example.}
	\label{fig:sigmoid_all}
\end{figure}
\begin{figure}[!htbp]
	\setlength{\abovecaptionskip}{0.0cm}
	\centering
	% 第一张图 (a)
	\subfloat[]{\includegraphics[width=0.9\textwidth]{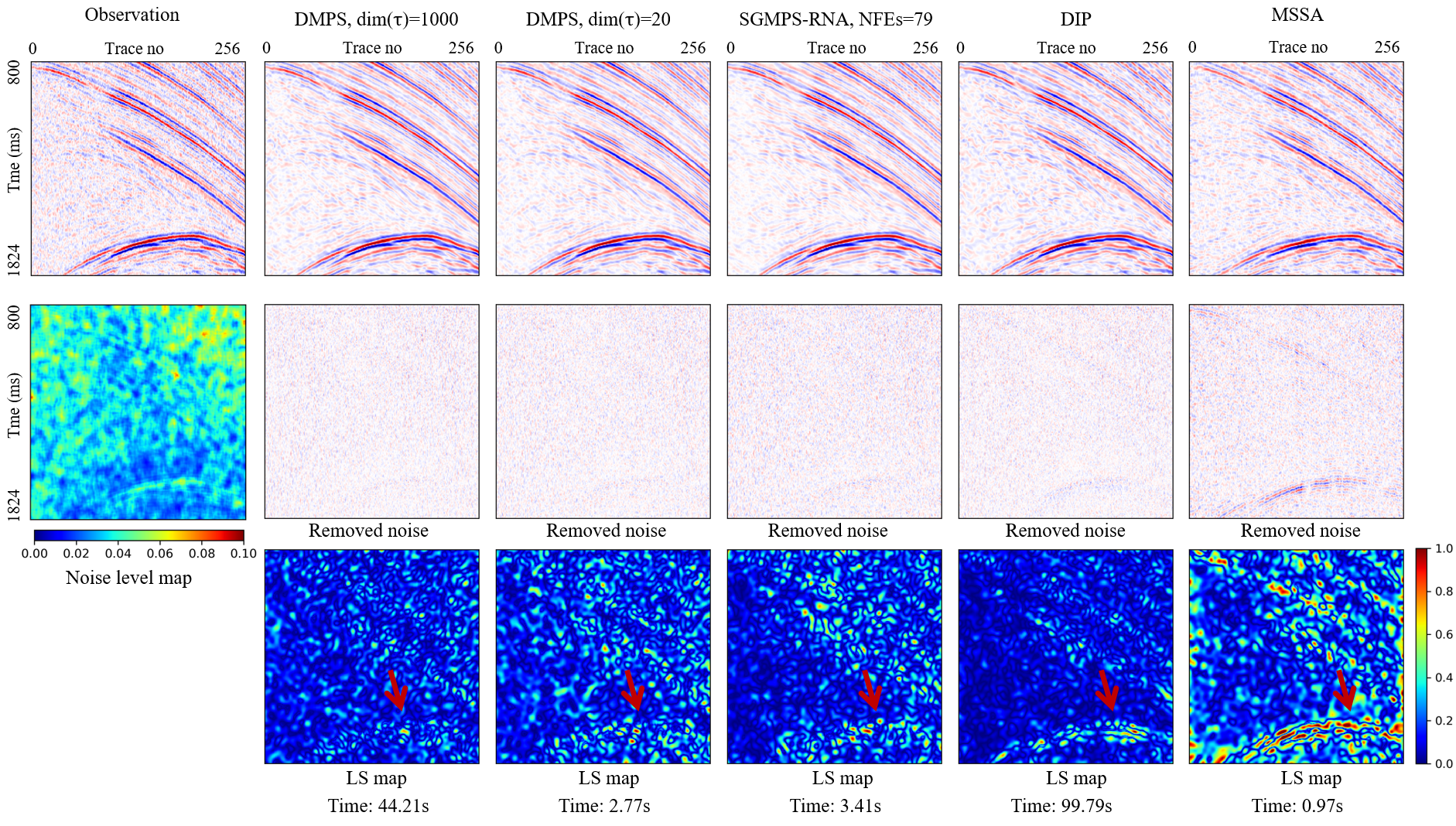}
		\label{fig:xj_dn}}\\[0.0cm]
	% 第二张图 (b)
	\subfloat[]{\includegraphics[width=0.9\textwidth]{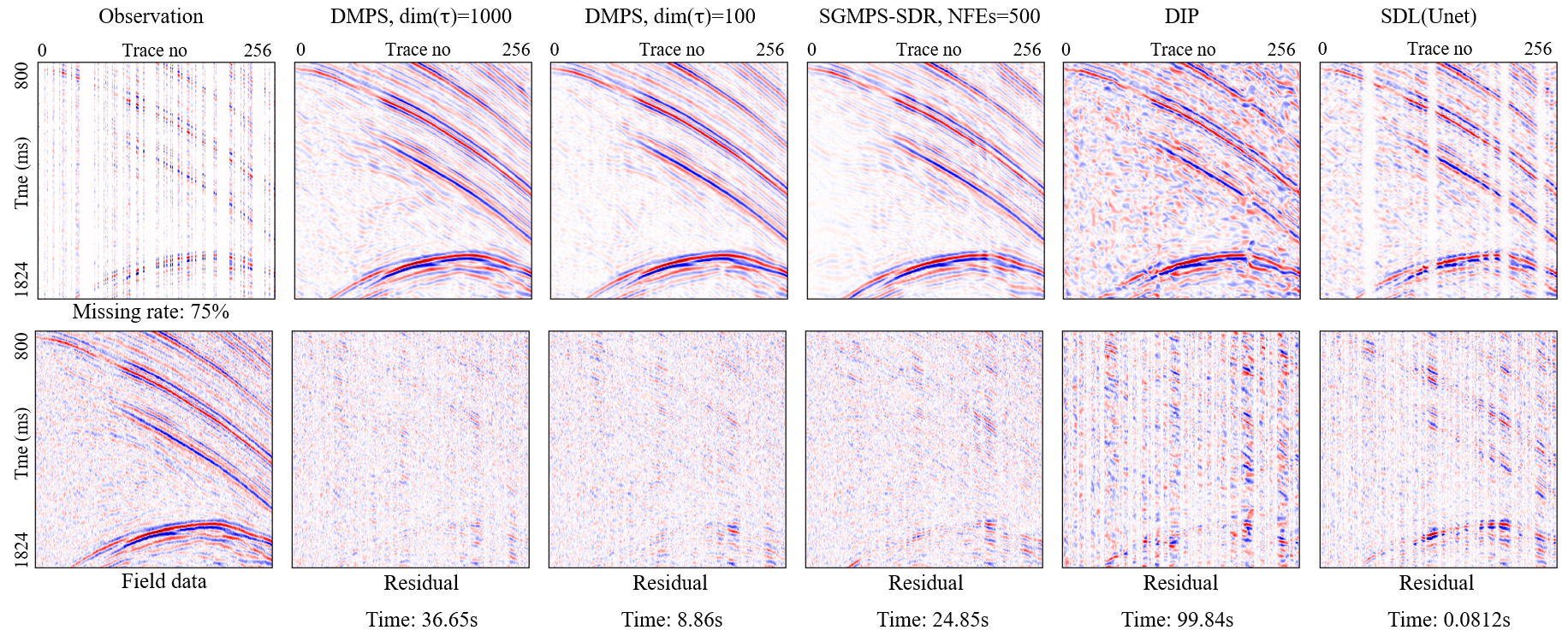}
		\label{fig:xj_int}}\\[0.0cm]
	% 第三张图 (c)
	\subfloat[]{\includegraphics[width=0.9\textwidth]{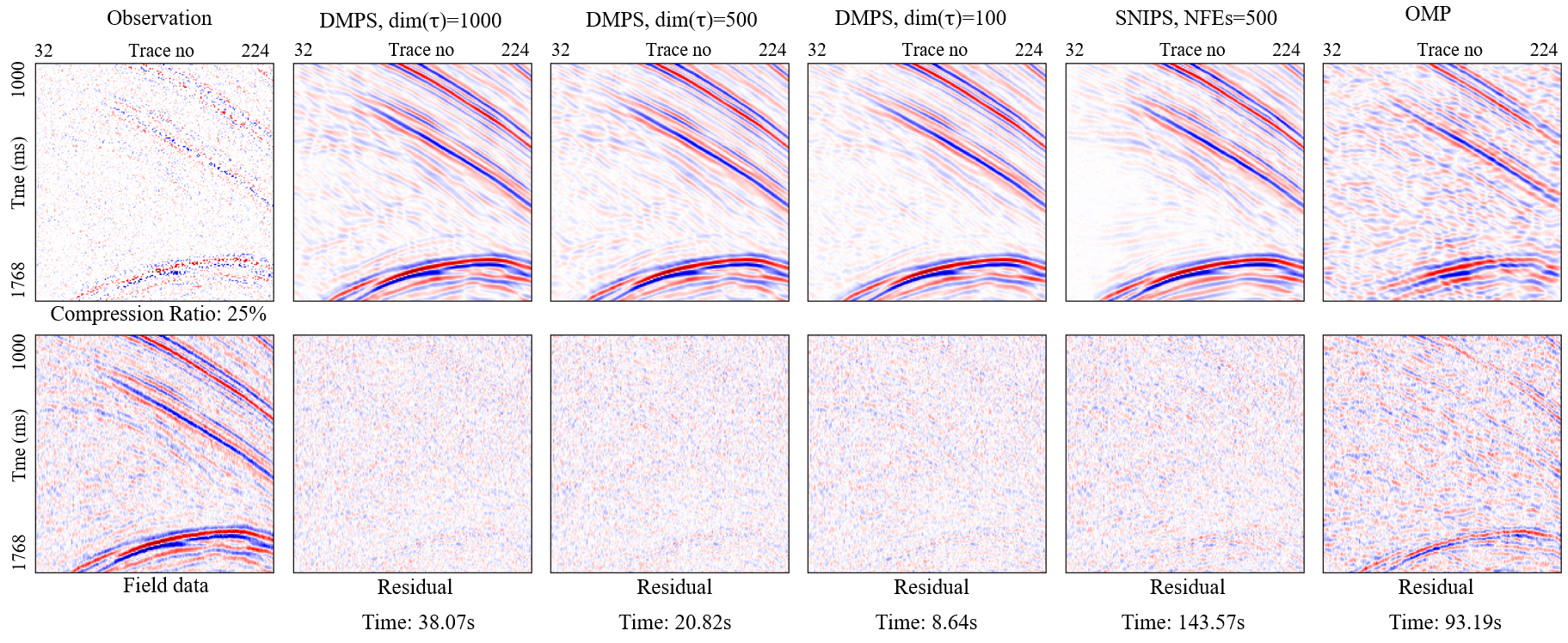}
		\label{fig:xj_cs}}
	\caption{Posterior sampling results of field data, taking tasks (a) denoising, (b) interpolation, and (c) compressed sensing as an example.}
	\label{fig:xj_all}
\end{figure}

\subsection{Multi-purpose posterior sampling for noisy inverse problems}

%\subsubsection{Synthetic Data Example}
%
%Field noise is much more complex than synthetic noise,
%
%\subsubsection{Field Data Example}

To quantitatively compare the sampling results across different inverse problems, we evaluate our method —referred to as Diffusion Model Posterior Sampling (DMPS)— on a complex synthetic model, Sigmoid (with shape $(192, 192)$, the time sampling interval is 2ms), which was not seen during the training of the generative models and thus represents an out-of-distribution (OOD) posterior sampling scenario. For each inverse problem, we compare our approach against competitive task-specific baseline methods.

For the denoising task, we compare DMPS with the following baselines: posterior sampling using a score-based generative model for random noise attenuation\cite{meng2025posterior} (abbreviated as SGMPS-RNA), Deep Image Prior (DIP)\cite{ulyanov2018deep}, and Multichannel Singular Spectrum Analysis (MSSA)\cite{oropeza2011simultaneous}. The SGM used in SGMPS-RNA is the open-source model released by \cite{meng2024generative}. In Figure~\ref{fig:sigmoid_dn}, the maximum absolute value of the clean data is 0.2, and the true noise level of the observation is set to $0.5 \times 0.2 = 0.1$. The NFEs for SGMPS-RNA depends on the noise level. When using the $\sigma(\mathbf{y})=0.1045$ predicted by VI-non-IID, the corresponding NFE is 146. The results of DIP and MSSA are selected from multiple runs with varying parameters, choosing the best-performing ones. The number of iterations of DIP is set to 500, and MSSA retains the nine largest singular values.
As shown in Figure~\ref{fig:sigmoid_dn}, compared to SGMPS-RNA, DIP, and MSSA, DMPS demonstrates superior performance in recovering complex geological structures, achieving higher SNR and SSIM values, especially when using the full trajectory (NFEs = 1000), and still maintaining competitive results under accelerated settings (NFEs = 20).
For the interpolation task, we compare against several baselines: posterior sampling using a score-based generative model for simultaneous denoising and reconstruction (abbreviated as SGMPS-SDR)\cite{meng2024stochastic}, DIP, and a supervised deep learning method using a U-Net \cite{ronneberger2015u}, denoted as SDL(Unet). SGMPS-SDR is also based on the publicly available SGM released by \cite{meng2024generative}, using a fixed number of NFEs = 500 without acceleration support.
As shown in Figure~\ref{fig:sigmoid_int}, the data suffer from 75\% regularly missing traces and additional noise at a level of $0.1 \times 0.2$. DDPM-based posterior sampling (DMPS) delivers the best results across both full and accelerated sampling (see red boxes), with minimal reconstruction residuals.
For the compressed sensing task, we compare our method against two baselines: Solving Noisy Inverse Problems Stochastically (SNIPS)\cite{kawar2021snips} and Orthogonal Matching Pursuit (OMP)\cite{tropp2007signal}. SNIPS is a score-based posterior sampling approach designed for a broad class of linear inverse problems. SNIPS requires an SGM trained on seismic data; we use the pretrained model released by \cite{meng2024generative}. It requires 500 function evaluations (NFEs), lacks an acceleration mechanism, and incurs high space complexity of $\mathcal{O}(d^2)$. Table \ref{tab:complexity} presents the time and space complexities of sampling for different generative models, where our method has the lowest complexity.
For DMPS, since the Fast Walsh–Hadamard Transform only supports data sizes that are powers of two, we adopt the Discrete Cosine Transform (DCT)~\cite{ahmed2006discrete} as the transformation matrix $\mathbf{W}$. For OMP, the dictionary matrix $\mathbf{D}$ is constructed using DCT, while the sensing matrix $\boldsymbol{\Phi} \in \mathbb{R}^{d_0 \times d}$ is a normalized random Gaussian matrix. The observation model is given by
$\mathbf{y} = \boldsymbol{\Phi} \mathbf{x}$.
Sparse coefficients are then recovered based on the measurement matrix $\mathbf{A} = \boldsymbol{\Phi} \mathbf{D}$ and the observation $\mathbf{y}$, with the sparsity level fixed at 300.
As shown in Figure~\ref{fig:sigmoid_cs}, the observation is acquired with a compression ratio of 0.25 and contains additive noise with a standard deviation of $0.1 \times 0.2$. DDPM achieves better results than both SNIPS (NFEs = 500) and OMP across different sampling settings (NFEs = 1000, 500, and 200), with the lowest reconstruction residuals.

\begin{table}[htbp]
	\centering
	\caption{Time and space complexity of different methods ($d$ is data dimension, $T$ is total number of steps).}
	\label{tab:complexity}
	\begin{tabular}{lcccc}
		\toprule
		Methods & DMPS & SGMPS-RNA & SGMPS-SDR & SNIPS\\
		\midrule
		Time complexity & $\mathcal{O}(\text{dim}(\tau) \cdot d)$, & $\mathcal{O}(L_n \cdot d)$, & $\mathcal{O}(T \cdot d)$ & $\mathcal{O}(T \cdot d)$ \\
		& $\text{dim}(\tau) < T$ & $L_n < T$ &  &  \\
		%		\midrule
		Space complexity & $\mathcal{O}(d)$ & $\mathcal{O}(d)$ & $\mathcal{O}(d)$ & $\mathcal{O}(d^2)$ \\
		\bottomrule
	\end{tabular}
\end{table}

We further evaluate posterior sampling results for  three tasks on field data (with shape of $(256, 256)$, the time sampling interval is 4ms), which represents an OOD generalization scenario. As shown in Figure~\ref{fig:xj_all}, this real dataset exhibits large spatial variations in both amplitude and pointwise noise levels. The number of iterations of DIP is set to 3000, and MSSA retains the nine largest singular values.
In the denoising task (Figure ~\ref{fig:xj_dn}), DMPS effectively suppresses noise with minimal leakage of useful signals and is among the fastest methods, second only to MSSA in runtime.
In the interpolation task (Figure ~\ref{fig:xj_int}), DMPS achieves the smallest reconstruction residuals while also suppressing noise. Remarkably, it reaches performance close to that of 1000 sampling steps using only 20 steps. It is also the fastest method apart from the supervised deep learning approach.
In the compressed sensing task (Figure ~\ref{fig:xj_cs}), DMPS achieves better reconstruction than SNIPS (NFEs = 500) and OMP with only 200 steps, while being the most efficient in terms of runtime.

\begin{figure}[!htbp]
%	\vspace{-2mm}
	\centering
	\includegraphics[width=\textwidth]{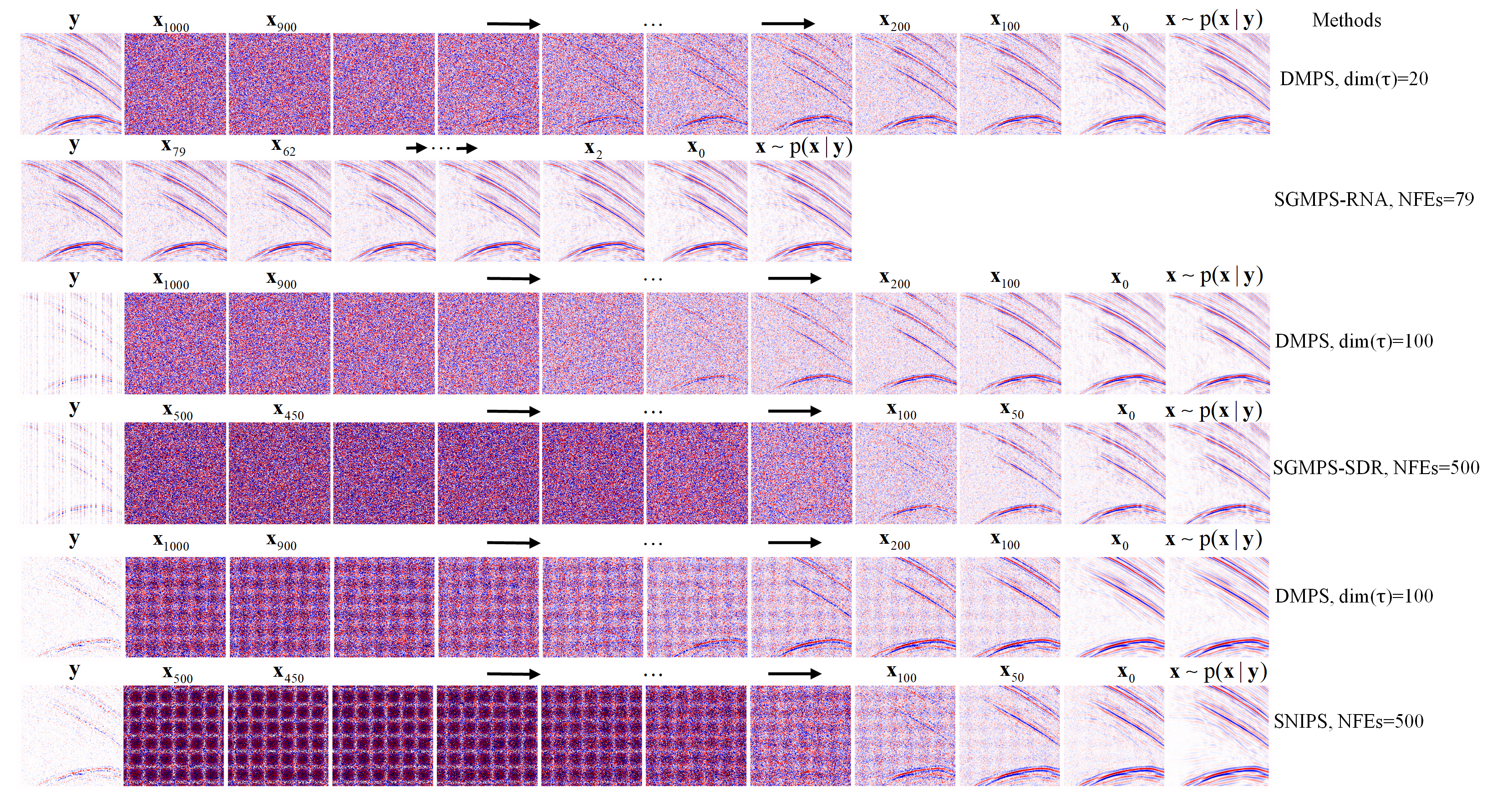}
	\caption{Schematic diagram of the posterior sampling trajectory of different models in Figure \ref{fig:xj_all}.}
	\label{fig:XJ_PS}
%	\vspace{-2mm}
\end{figure}

Figure ~\ref{fig:XJ_PS} illustrates the sampling trajectories of all generative posterior sampling methods used in Figure ~\ref{fig:xj_all}. The sampling trajectory of DDPM corresponds to a subsequence sampling path, whereas the other methods use full-sequence sampling. For clarity, each trajectory displays only a few intermediate variables selected at regular intervals.

\section{Discussion} 
This paper presents a diffusion model-based posterior sampling method suitable for solving linear inverse problems. The method can also be extended to posterior sampling for nonlinear inverse problems. In all cases, generative modeling of seismic data to obtain learnable parameters for diffusion models is an essential step.
There are several ways, such as variational methods or Bayes' theorem, to design posterior sampling algorithms based on a pre-trained diffusion model without relying on task-specific conditional models.
In this paper, the conditional model \( p_\theta(\mathbf{x}_{0:T}, \mathbf{y}) \) is formulated via variational inference, using the same training objective as the unconditional DDPM, \( p_\theta(\mathbf{x}_{0:T}) \), thus eliminating the need for retraining. 
Additionally, based on Bayes' theorem, posterior sampling can be performed by constructing a conditional score function $\nabla_{\mathbf{x}_t} \log p(\mathbf{x}_t \mid \mathbf{y})$—i.e., the gradient of the log conditional distribution~\cite{chung2023diffusion, meng2024stochastic}, using a pre-trained diffusion model, which we consider a promising direction for future research.

\section{Conclusion} 
In this paper, we propose a diffusion model enhanced with a novel power noise schedule for generative modeling of seismic data, and introduce a non-Markovian sampling mechanism to enable fast and quality-controllable unconditional generation. We further present a fast and efficient unsupervised posterior sampling method for various noisy inverse problems using the trained unconditional diffusion model, eliminating the need for task-specific retraining. Unlike discriminative approaches, our method can generate multiple high-quality stochastic solutions, allowing users to interactively perform posterior sampling under different signal-to-noise ratio conditions. Experiments on unconditional generation and posterior sampling across different tasks demonstrate the superiority of our approach in seismic data generation, accelerated sampling, task-adaptive posterior sampling, and out-of-distribution generalization.

%\section{Acknowledgements} 
%This work was supported in part by
%the National Key $\&$ Program of China under Grant 2020YFA0713403 and 2020YFA0713400, in part by the China National Postdoctoral Program for Innovative Talents
%under Grant BX20230284, in part by the Postdoctoral Research Project of Shaanxi Province.

\bibliographystyle{unsrt}  
\bibliography{ref}  %%% Remove comment to use the external .bib file (using bibtex).

\begin{thebibliography}{10}

\bibitem{yu2019deep}
S.~Yu, J.~Ma, and W.~Wang.
\newblock Deep learning for denoising.
\newblock {\em Geophysics}, 84(6):1--107, 2019.

\bibitem{meng2023learning}
Chuangji Meng, Jinghuai Gao, Yajun Tian, and Zhen Li.
\newblock Learning to decouple and generate seismic random noise via invertible
  neural network.
\newblock {\em IEEE Transactions on Geoscience and Remote Sensing}, 61:1--16,
  2023.

\bibitem{wang2019deep}
Benfeng Wang, Ning Zhang, Wenkai Lu, and Jialin Wang.
\newblock Deep-learning-based seismic data interpolation: A preliminary result.
\newblock {\em Geophysics}, 84(1):V11--V20, 2019.

\bibitem{kaur2021seismic}
Harpreet Kaur, Nam Pham, and Sergey Fomel.
\newblock Seismic data interpolation using deep learning with generative
  adversarial networks.
\newblock {\em Geophysical Prospecting}, 69(2):307--326, 2021.

\bibitem{araya2018deep}
Mauricio Araya-Polo, Joseph Jennings, Amir Adler, and Taylor Dahlke.
\newblock Deep-learning tomography.
\newblock {\em The Leading Edge}, 37(1):58--66, 2018.

\bibitem{zhang2019seismic}
Mi~Zhang, Yang Liu, Min Bai, and Yangkang Chen.
\newblock Seismic noise attenuation using unsupervised sparse feature learning.
\newblock {\em IEEE Transactions on Geoscience and Remote Sensing},
  57(12):9709--9723, 2019.

\bibitem{yang2023porosity}
Liuqing Yang, Sergey Fomel, Shoudong Wang, Xiaohong Chen, Wei Chen, Omar~M
  Saad, and Yangkang Chen.
\newblock Porosity and permeability prediction using a transformer and periodic
  long short-term network.
\newblock {\em Geophysics}, 88(1):WA293--WA308, 2023.

\bibitem{tian2023frequency}
Yajun Tian, Alexey Stovas, Jinghuai Gao, Chuangji Meng, and Chun Yang.
\newblock Frequency-dependent avo inversion and application on tight sandstone
  gas reservoir prediction using deep neural network.
\newblock {\em IEEE Transactions on Geoscience and Remote Sensing}, 61:1--13,
  2023.

\bibitem{xiong2018seismic}
Wei Xiong, Xu~Ji, Yue Ma, Yuxiang Wang, Nasher~M AlBinHassan, Mustafa~N Ali,
  and Yi~Luo.
\newblock Seismic fault detection with convolutional neural network.
\newblock {\em Geophysics}, 83(5):O97--O103, 2018.

\bibitem{wu2020building}
Xinming Wu, Zhicheng Geng, Yunzhi Shi, Nam Pham, Sergey Fomel, and Guillaume
  Caumon.
\newblock Building realistic structure models to train convolutional neural
  networks for seismic structural interpretation.
\newblock {\em Geophysics}, 85(4):WA27--WA39, 2020.

\bibitem{kawar2021snips}
Bahjat Kawar, Gregory Vaksman, and Michael Elad.
\newblock Snips: Solving noisy inverse problems stochastically.
\newblock {\em Advances in Neural Information Processing Systems},
  34:21757--21769, 2021.

\bibitem{song2020score}
Yang Song, Jascha Sohl-Dickstein, Diederik~P Kingma, Abhishek Kumar, Stefano
  Ermon, and Ben Poole.
\newblock Score-based generative modeling through stochastic differential
  equations.
\newblock {\em arXiv preprint arXiv:2011.13456}, 2020.

\bibitem{song2019generative}
Yang Song and Stefano Ermon.
\newblock Generative modeling by estimating gradients of the data distribution.
\newblock {\em Advances in neural information processing systems},
  32:11895--11907, 2019.

\bibitem{ho2020denoising}
Jonathan Ho, Ajay Jain, and Pieter Abbeel.
\newblock Denoising diffusion probabilistic models.
\newblock {\em Advances in neural information processing systems},
  33:6840--6851, 2020.

\bibitem{meng2024generative}
Chuangji Meng, Jinghuai Gao, Yajun Tian, Hongling Chen, and Renyu Luo.
\newblock Generative modeling of seismic data using score-based generative
  models.
\newblock In {\em 85th EAGE Annual Conference \& Exhibition (including the
  Workshop Programme)}, volume 2024, pages 1--5. European Association of
  Geoscientists \& Engineers, 2024.

\bibitem{meng2025generative}
C~Meng, J~Gao, Y~Tian, H~Chen, L~Zhou, J~Chen, Q~Du, and Y~Li.
\newblock Generative modeling of seismic data using diffusion models and its
  application to multi-purpose seismic inverse problems.
\newblock In {\em 86th EAGE Annual Conference \& Exhibition}, volume 2025,
  pages 1--5. European Association of Geoscientists \& Engineers, 2025.

\bibitem{wang2024controllable}
Fu~Wang, Xinquan Huang, and Tariq Alkhalifah.
\newblock Controllable seismic velocity synthesis using generative diffusion
  models.
\newblock {\em Journal of Geophysical Research: Machine Learning and
  Computation}, 1(3):e2024JH000153, 2024.

\bibitem{feng2024analysis}
Qiankun Feng, Shigang Wang, and Yue Li.
\newblock Analysis of das seismic noise generation and elimination process
  based on mean-sde diffusion model.
\newblock {\em IEEE Transactions on Geoscience and Remote Sensing}, 62:1--13,
  2024.

\bibitem{peng2024seismic}
Junheng Peng, Yong Li, Zhangquan Liao, Xuben Wang, and Xingyu Yang.
\newblock Seismic data strong noise attenuation based on diffusion model and
  principal component analysis.
\newblock {\em IEEE Transactions on Geoscience and Remote Sensing}, 62:1--11,
  2024.

\bibitem{meng2025posterior}
Chuangji Meng, Jinghuai Gao, Baohai Wu, Hongling Chen, and Yajun Tian.
\newblock Posterior sampling for random noise attenuation via score-based
  generative models.
\newblock {\em Geophysics}, 90(2):V83--V95, 2025.

\bibitem{peng2025fast}
Junheng Peng, Yong Li, Yingtian Liu, Mingwei Wang, Zhangquan Liao, and Xiaowen
  Wang.
\newblock Fast diffusion model for seismic data noise attenuation.
\newblock {\em Geophysics}, 90(4):1--55, 2025.

\bibitem{liu2024generative}
Qi~Liu and Jianwei Ma.
\newblock Generative interpolation via a diffusion probabilistic model.
\newblock {\em Geophysics}, 89(1):V65--V85, 2024.

\bibitem{wei2024seismic}
Xiaoli Wei, Chunxia Zhang, Hongtao Wang, Chengli Tan, Deng Xiong, Baisong
  Jiang, Jiangshe Zhang, and Sang-Woon Kim.
\newblock Seismic data interpolation via denoising diffusion implicit models
  with coherence-corrected resampling.
\newblock {\em IEEE Transactions on Geoscience and Remote Sensing}, 62:1--17,
  2024.

\bibitem{meng2024stochastic}
Chuangji Meng, Jinghuai Gao, Yajun Tian, Hongling Chen, Wei Zhang, and Renyu
  Luo.
\newblock Stochastic solutions for simultaneous seismic data denoising and
  reconstruction via score-based generative models.
\newblock {\em IEEE Transactions on Geoscience and Remote Sensing}, 2024.

\bibitem{wang2024seisfusion}
Shuang Wang, Fei Deng, Peifan Jiang, Zishan Gong, Xiaolin Wei, and Yuqing Wang.
\newblock Seisfusion: Constrained diffusion model with input guidance for 3d
  seismic data interpolation and reconstruction.
\newblock {\em IEEE Transactions on Geoscience and Remote Sensing}, 2024.

\bibitem{shi2024generative}
Xingchen Shi, Shijun Cheng, Weijian Mao, and Wei Ouyang.
\newblock Generative diffusion model for seismic imaging improvement of
  sparsely acquired data and uncertainty quantification.
\newblock {\em IEEE Transactions on Geoscience and Remote Sensing}, 2024.

\bibitem{zhang2024seisresodiff}
Hao-Ran Zhang, Yang Liu, Yu-Hang Sun, and Gui Chen.
\newblock Seisresodiff: Seismic resolution enhancement based on a diffusion
  model.
\newblock {\em Petroleum Science}, 21(5):3166--3188, 2024.

\bibitem{zhu2023diffusion}
Donglin Zhu, Lei Fu, Vladimir Kazei, and Weichang Li.
\newblock Diffusion model for das-vsp data denoising.
\newblock {\em Sensors}, 23(20):8619, 2023.

\bibitem{zhang2024conditional}
Hao Zhang, Yuanyuan Li, and Jianping Huang.
\newblock Conditional denoising diffusion probabilistic model for seismic
  diffraction separation and imaging.
\newblock {\em IEEE Transactions on Geoscience and Remote Sensing}, 2024.

\bibitem{durall2023deep}
Ricard Durall, Ammar Ghanim, Mario~Ruben Fernandez, Norman Ettrich, and Janis
  Keuper.
\newblock Deep diffusion models for seismic processing.
\newblock {\em Computers \& Geosciences}, 177:105377, 2023.

\bibitem{chen2025unsupervised}
Hongling Chen, Jie Chen, Mauricio Sacchi, Jinghuai Gao, and Ping Yang.
\newblock Unsupervised seismic acoustic impedance inversion based on generative
  diffusion model.
\newblock {\em Geophysics}, 90(4):1--98, 2025.

\bibitem{baldassari2024conditional}
Lorenzo Baldassari, Ali Siahkoohi, Josselin Garnier, Knut Solna, and Maarten~V
  de~Hoop.
\newblock Conditional score-based diffusion models for bayesian inference in
  infinite dimensions.
\newblock {\em Advances in Neural Information Processing Systems}, 36, 2024.

\bibitem{cheng2025agenerative}
Shijun Cheng, Randy Harsuko, and Tariq Alkhalifah.
\newblock A generative foundation model for an all-in-one seismic processing
  framework.
\newblock {\em arXiv preprint arXiv:2502.01111}, 2025.

\bibitem{song2020denoising}
Jiaming Song, Chenlin Meng, and Stefano Ermon.
\newblock Denoising diffusion implicit models.
\newblock {\em arXiv preprint arXiv:2010.02502}, 2020.

\bibitem{nichol2021improved}
Alexander~Quinn Nichol and Prafulla Dhariwal.
\newblock Improved denoising diffusion probabilistic models.
\newblock In {\em International conference on machine learning}, pages
  8162--8171. PMLR, 2021.

\bibitem{kawar2022denoising}
Bahjat Kawar, Michael Elad, Stefano Ermon, and Jiaming Song.
\newblock Denoising diffusion restoration models.
\newblock {\em Advances in Neural Information Processing Systems},
  35:23593--23606, 2022.

\bibitem{kadkhodaie2021stochastic}
{Kadkhodaie, Zahra and Simoncelli, Eero}.
\newblock Stochastic solutions for linear inverse problems using the prior
  implicit in a denoiser.
\newblock {\em Advances in Neural Information Processing Systems},
  34:13242--13254, 2021.

\bibitem{chung2023diffusion}
Hyungjin Chung, Jeongsol Kim, Michael~Thompson Mccann, Marc~Louis Klasky, and
  Jong~Chul Ye.
\newblock Diffusion posterior sampling for general noisy inverse problems.
\newblock In {\em The Eleventh International Conference on Learning
  Representations}, 2023.

\bibitem{meng2022seismic}
Chuangji Meng, Jinghuai Gao, Yajun Tian, and Zhiqiang Wang.
\newblock Seismic random noise attenuation based on non-iid pixel-wise gaussian
  noise modeling.
\newblock {\em IEEE Transactions on Geoscience and Remote Sensing}, 60:1--16,
  2022.

\bibitem{song2020improved}
{Song, Yang and Ermon, Stefano}.
\newblock Improved techniques for training score-based generative models.
\newblock {\em Advances in neural information processing systems},
  33:12438--12448, 2020.

\bibitem{fino1976unified}
Fino and Algazi.
\newblock Unified matrix treatment of the fast walsh-hadamard transform.
\newblock {\em IEEE Transactions on Computers}, 100(11):1142--1146, 1976.

\bibitem{ulyanov2018deep}
Dmitry Ulyanov, Andrea Vedaldi, and Victor Lempitsky.
\newblock Deep image prior.
\newblock In {\em Proceedings of the IEEE conference on computer vision and
  pattern recognition}, pages 9446--9454, 2018.

\bibitem{oropeza2011simultaneous}
Vicente Oropeza and Mauricio Sacchi.
\newblock Simultaneous seismic data denoising and reconstruction via
  multichannel singular spectrum analysis.
\newblock {\em Geophysics}, 76(3):V25--V32, 2011.

\bibitem{ronneberger2015u}
Olaf Ronneberger, Philipp Fischer, and Thomas Brox.
\newblock U-net: Convolutional networks for biomedical image segmentation.
\newblock In {\em International Conference on Medical image computing and
  computer-assisted intervention}, pages 234--241. Springer, 2015.

\bibitem{tropp2007signal}
Joel~A Tropp and Anna~C Gilbert.
\newblock Signal recovery from random measurements via orthogonal matching
  pursuit.
\newblock {\em IEEE Transactions on information theory}, 53(12):4655--4666,
  2007.

\bibitem{ahmed2006discrete}
Nasir Ahmed, T\_ Natarajan, and Kamisetty~R Rao.
\newblock Discrete cosine transform.
\newblock {\em IEEE transactions on Computers}, 100(1):90--93, 2006.

\end{thebibliography}

\end{document}